\documentclass[twocolumn,superscriptaddress, reprint,aps]{revtex4-1}

\usepackage{graphicx,amsmath,amssymb,todonotes, mathtools, appendix}	
\usepackage{hyperref}
\usepackage{epstopdf}

  \newcommand{\computingLab}{University of Oxford, Department of Computer Science, Wolfson Building, Parks Road, Oxford, OX1 3QD, UK}
  \newcommand{\clarendonLab}{University of Oxford, Department of Physics, Clarendon Laboratory, Parks Road, Oxford, OX1 3PU, UK}
  \newcommand{\rudolfPeierls}{University of Oxford, Department of Physics, Rudolf Peierls Centre for Theoretical Physics, 1 Keble Road, Oxford, OX1 3NP, UK}
  \newcommand{\Imperial}{Imperial College, Department of Mathematics, 180 Queen's Gate, London, SW7 2AZ, UK }
  \newcommand{\york}{University of York, Department of Electronics, YO10 5DD, York, U.K.}

  \newcommand{\degreeC}{\,^{\circ}{\rm C}}
  \newcommand{\degreeCpMin}{\,^{\circ}{\rm C \mbox{ min}^{-1}}}      
  \newcommand*{\plimsoll}{{\ensuremath{-\kern-4pt{\ominus}\kern-4pt-}}}

\newcommand{\teoedit}[1]{{\color{black} #1}}

  \newcommand{\pictureWithCaption}[4]{ 
    \begin{figure*}[#1]
      \includegraphics[width=#2\textwidth]{#3}
      \caption{\label{#3} #4  }
    \end{figure*}
  }

  \newcommand{\pictureWithCaptionSingleColumn}[4]{ 
    \begin{figure}[#1]
      \centering
      \includegraphics[width=#2\columnwidth]{#3}
      \caption{#4 \label{#3} }
    \end{figure}
  }
  \newcommand{\gnuplotpicturesWithCaptionSingleColumn}[7]{ 

    \begin{figure}[#1]
      \centering
\vspace{-5mm}

      \includegraphics[width=#2\columnwidth, angle=-90]{#3}

	\vspace{-15mm}

\includegraphics[width=#2\columnwidth, angle=-90]{#4}

	\vspace{-15mm}

\includegraphics[width=#2\columnwidth, angle=-90]{#5}

	\vspace{-8mm}

      \caption{#6 \label{#7} }
    \end{figure}
  }

\newcommand{\staple}{p}    
\newcommand{\stateSpace}{{S}}
\newcommand{\state}{s}

\newcommand{\eqn}{Eq.\ }

  \newcommand{\stateA}{\mathsf{a}}
  \newcommand{\stateB}{\mathsf{b}}
  \newcommand{\stateC}{\mathsf{c}}
  \newcommand{\stateD}{\mathsf{d}}
  
  \newcommand{\molar}{\mbox{M}}
  
  \newcommand{\cf}{cf.\ }

\newcommand{\sybr}{\mbox{SYBR}^{\circledR}}
\newcommand{\face}{\mathsf{F}}
\newcommand{\faces}{\mathsf{F}}
\newcommand{\weight}{\mathsf{W}}
\newcommand{\rate}{\sigma}
\newcommand{\edge}{e}
\newcommand{\vertex}{v}
\newcommand{\edges}{\mathsf{E}}
\newcommand{\vertices}{\mathsf{V}}
\newcommand{\distance}{r}
\newcommand{\graph}{\mathsf{H}}

\newcommand{\status}{\mathsf{L}}

\begin{document}

\title{Modelling DNA Origami Self-Assembly at the Domain Level}

\author{Frits Dannenberg}
\affiliation{\computingLab}

\author{Katherine E. Dunn}
\affiliation{\clarendonLab}
\affiliation{\york}

\author{Jonathan Bath}
\affiliation{\clarendonLab}

\author{Marta Kwiatkowska} 
\affiliation{\computingLab}

\author{Andrew J. Turberfield}
\affiliation{\clarendonLab}

\author{Thomas E. Ouldridge}
\affiliation{\rudolfPeierls}
\affiliation{\Imperial}

\begin{abstract}	
	We present a modelling framework, and basic model parameterization, for the study of DNA origami folding at the level of DNA domains. Our approach is explicitly kinetic and does not assume a specific folding pathway. The binding of each staple  is associated with a free-energy change that depends on staple sequence, the possibility of coaxial stacking with neighbouring domains, and the entropic cost of constraining the scaffold by inserting staple crossovers. A rigorous thermodynamic model is difficult to implement as a result of the complex, multiply connected geometry of the scaffold: we present a solution to this problem for planar origami. Coaxial stacking of helices and entropic terms, particularly when loop closure exponents are taken to be larger than those for ideal chains, introduce interactions between staples. These cooperative interactions lead to the prediction of sharp assembly transitions with notable hysteresis that are consistent with experimental observations. We  show that the model reproduces the experimentally observed consequences of reducing staple concentration, accelerated cooling and absent staples. We also present a simpler methodology that gives consistent results and can be used to study a wider range of systems including non-planar origami.     	
\end{abstract}

\maketitle

\section{Introduction}

Recently, the exquisite specificity of Watson-Crick base pairing has been harnessed to create artificial nanoscale structures from DNA \cite{Chen91,Goodman2005,DNA-Rothemund-2006-0002,Zheng2009,Douglas09,Andersen09,Dietz2009,Han2011,Ke2012,Zhang2014}.  One of the most popular techniques for assembling DNA nanostructures is known as DNA origami. Pioneered by Rothemund \cite{DNA-Rothemund-2006-0002}, this approach involves the folding of a long ``scaffold" strand of DNA (often the genome of the M13 bacteriophage, approximately 7300 bases in length) by hybridization with a set of much shorter ``staple" strands. Most staples are designed to be complementary to two or more distinct parts of the scaffold; hybridization to staples causes the scaffold to fold into a target shape. DNA origami has been used to construct a wide variety of 2D and 3D structures \cite{DNA-Rothemund-2006-0002,Douglas09,Andersen09,Dietz2009,Han2011,Zhang2014}.  Origami nanostructures can be functionalized with sub-nanometre precision, allowing them to function as molecular breadboards for the construction of optical devices \cite{Kuzyk2012} and light-harvesting complexes \cite{Dutta2011}, tracks for decision-making robots \cite{Wickham2012,Tomov2013}, 	and scaffolds for studying enzymatic cascades \cite{Fu2012}. Origami structures can act as `frames', `rulers' and `handles' for single-molecule manipulation \cite{Endo2012, Pfitzner2013}  or harnesses for connecting multiple motor proteins \cite{Derr2012}. Other suggested applications include drug delivery \cite{Douglas2012,Amir2014}.

Although DNA origami has been remarkably successful, open questions about the assembly process remain. Folding yield is variable, and even when a structure is largely folded missing staples could affect its mechanical properties  \cite{Chen2014} and addressability. An understanding of general principles for optimizing yield and folding rate would allow the improvement of current designs and the assembly of larger and more complex structures.  A number of recent experiments have probed the details of origami folding \cite{Song2012,Sobczak2012,Martin2012,Ke2012b,Arbona2013,Wei2013,Wei2014}. Assembly is cooperative \cite{Sobczak2012,Arbona2013}, occurs across a narrow temperature window \cite{Sobczak2012}, exhibits hysteresis \cite{Sobczak2012,Wei2013,Arbona2013} and is  highly sensitive to staple design \cite{Ke2012b, Martin2012,Arbona2013}.  
    
Current theoretical modelling is very limited. Although the thermodynamics of isolated duplex formation are well understood and quantified  \cite{SantaLucia2004}, it is not possible to model origami formation as a sequence of independent staple-binding events without losing much of the underlying physics. In particular, such an approach cannot describe cooperative interactions whereby the presence of a bound staple affects the binding of other staples to the scaffold.  Recent evidence suggests that cooperative interactions affect origami folding significantly \cite{Sobczak2012,Wei2013,Arbona2013}, leading to sharp formation transitions as the temperature is lowered and contributing to hysteresis in heating and cooling experiments.    
     
The  physical causes of cooperativity during origami folding are subtle. One might hope to draw analogies with the nucleation of crystals: until a crystal exceeds a threshold size, the lower free energy of the crystalline phase is offset by a substantial interfacial cost. Theoretical work has suggested that nucleated growth does occur in another class of DNA nanostructures that assemble from many short strands (``bricks") without a guiding scaffold \cite{Reinhardt2014}. In this case, cooperative, nucleated assembly is expected: a high proportion of the bricks of a pre-nucleated structure are only bound to one other brick, whereas bricks binding to a larger structure can form two or more bonds.        
For DNA origami, however, the analogy with crystal nucleation is not so obvious. Staples bind to the scaffold only, not to other staples, and can hybridize to the same number of scaffold domains at any stage of assembly. Arbona {\it et al.} \cite{Arbona2013,Song2012} have presented a model of origami folding in which cooperativity arises from two effects. 
\begin{itemize}
	\item A bound staple generally holds two or more distinct parts of the longer scaffold strand in close proximity, forming scaffold loops and incurring an entropic penalty. Arbona {\it et al.} postulate that the binding of one staple to the origami might facilitate the binding of a second by shortening the loop which it must enclose \cite{Arbona2013}.
	\item Arbona {\it et al.} include a phenomenological term through which the density of nearby staples is taken to stabilize the binding of an additional staple to the scaffold; they suggest that attractive interactions between strands mediated by divalent cations might be responsible for such an interaction \cite{Arbona2013}. 
\end{itemize} 
Although an important contribution, there are some drawbacks to their approach.  For the sake of tractability, staple binding is assumed to follow a single, well-defined pathway.  
This assumption means that the actual staple binding probabilities in the steady state are not consistent with the free energies used as inputs to the model; the consequences of this discrepancy are not explored. There is no timescale in the model, so it is unclear how properties such as apparent hysteresis should be interpreted. 
  
In this work we demonstrate a general  `global' model for the folding of any origami that has a structure that can be represented as a planar graph. The global model is thermodynamically self-consistent, has  explicit kinetics and allows for the analysis of competing folding pathways.  The model naturally generates  cooperative effects in staple binding: cooperativity arises as a result of the influence of bound staples on the proximity of the binding domains of other staples and from coaxial stacking of duplexes.  
We explore changes in origami folding as the strength of the cooperativity factors are varied, and demonstrate that the model is consistent with experimental observations of the consequences of reducing staple concentration, accelerated cooling and absent staples. We also present a simplified `local' model which can be extended to non-planar graphs. The local approach is not a true thermodynamic model, but it does give similar results to the global model and  deviations are  systematic and explicable.


\teoedit{We recently used the local model to predict folding trajectories in an unusual origami with a scaffold composed of two identical halves  \cite{Dunn2015}. In this system, two of each staple type can bind to a single scaffold in two possible configurations, leading to a vast number of possible hybridized structures among which are a number of approximately degenerate but geometrically distinct well-formed assembly products. The distribution between well-formed products is strongly dependent on cooperative effects in staple binding, and the model is remarkably successful in predicting the effects of rationally engineered changes in staple-scaffold interactions.}
 
\pictureWithCaptionSingleColumn{t}{0.99}{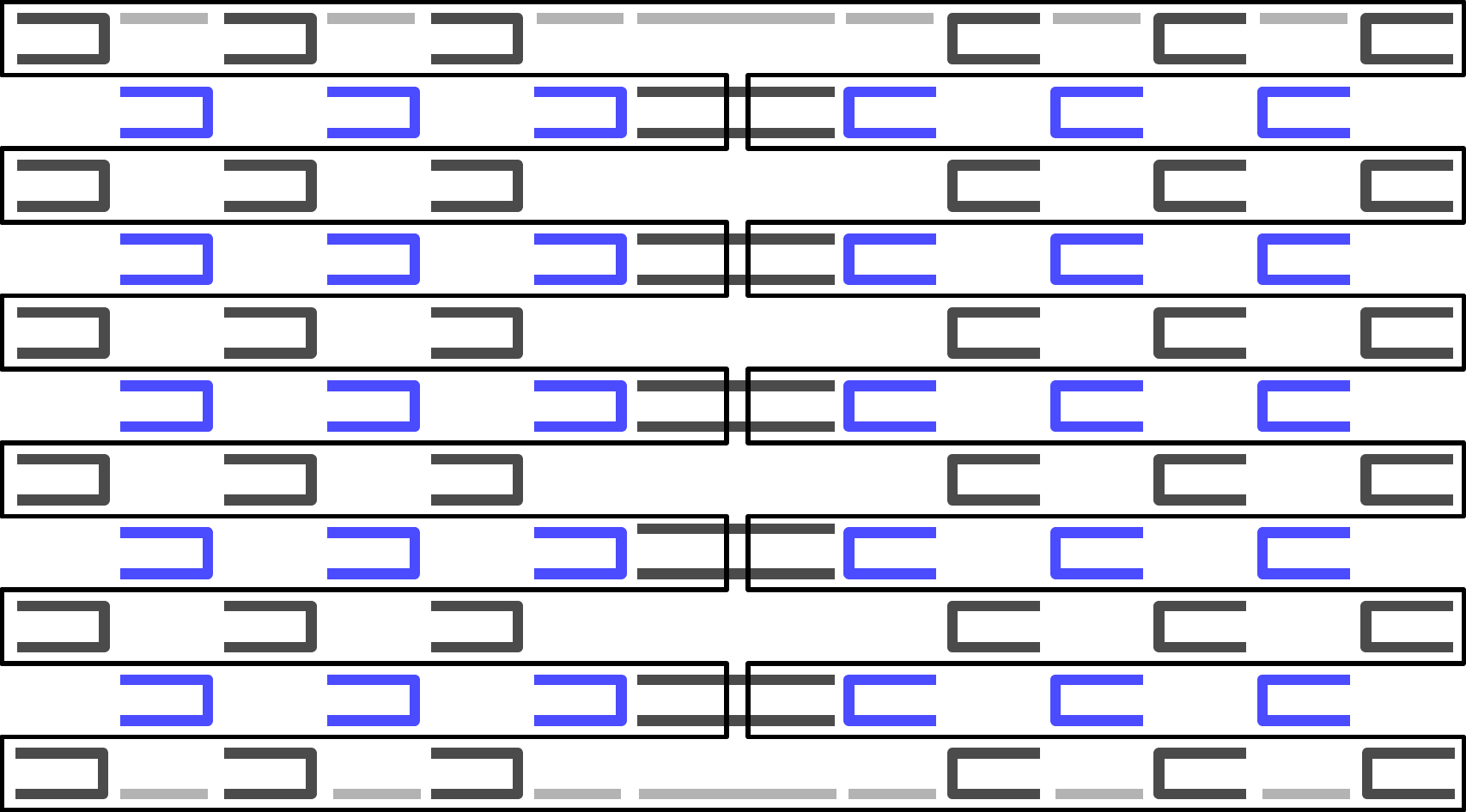}{
	A simple origami design  studied here. Two-domain staples are shown in blue and black, and single-domain staples in grey. The scaffold, shown as a thin black line, is a long loop of DNA that hybridizes to each domain of each staple. A specific implementation of this design is described in Ref. \cite{Dunn2015}.
	\label{layout}
}

\pictureWithCaptionSingleColumn{t}{0.99}{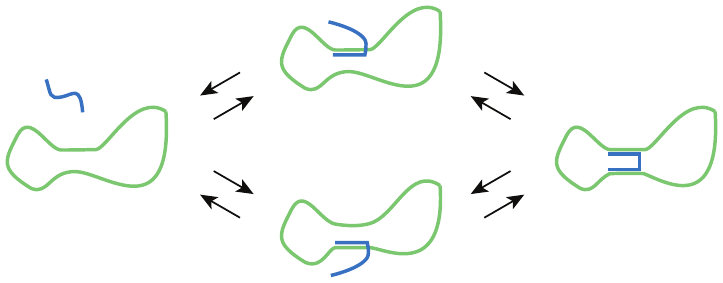}{
	Transitions in the model. Four possible bonding  configurations of two domains of a scaffold which can bind to staple $i$ are shown. From left to right these configurations are denoted $p_i=00$, $p_i=01, 10$ (half-bound) and $p_i=11$ (fully bound). A fifth state with two half-bound staples ($p_i = 12$) is not shown.  
}

\section{Model and Methods}\label{the model}
\subsection{Model state space} 
 
We model the folding of an isolated scaffold surrounded by a large excess of staples and describe the folding of this origami at the level of \emph{domains}. Such an abstraction is common in models of complex nucleic acid nanostructures and strand displacement systems \cite{Phillips2009,Arbona2013,Chen2013,Reinhardt2014}.
A domain is a contiguous series of bases, typically 5 to 20 nucleotides long, that is designed \teoedit{through its base sequence} to bind \teoedit{selectively} to a complementary domain as a cooperative unit. 
Origami scaffolds typically have a few hundred domains whereas staples typically have between one and five.

For simplicity, we will restrict  examples in this work to staples with at most two domains (Fig. \ref{layout}), although our approach can be generalized to staples with more domains (see 
Appendix \ref{multidomainstaples}).
We describe a two-domain staple as half-bound if only one domain is hybridized to the scaffold and fully bound if both domains are bound (Fig. \ref{scaffold}).
We assume that only fully complementary domains can hybridize, ignoring weaker interactions that result from inevitable partial sequence complementarity between other pairs of domains.
These interactions could be included, but at a high price, as this would significantly expand the state space of the model. We note that it would be an even harder problem to consider all possible interactions that are not consistent with the abstraction to domains, including hybridization of a staple to more than one scaffold domain and the formation of secondary structure within the scaffold itself.  

Let the design consist of $k$ staples and let $p_{i}$ denote the bonding configuration of the scaffold domains which interact with the $i$-th staple.
For single-domain staples we define $\staple_{i} \in \{ 0,1 \}  $ where
\begin{itemize}
	\item 0: a staple is not bound to the scaffold domain;
	\item 1: a staple is bound to the scaffold domain.
\end{itemize}
For two-domain staples we have $\staple_{i} \in  \{ 00, 10, 01, 11, 12 \} $ where:
\begin{itemize}
	\item	00: no staple bound to either scaffold domain;
	\item	10: a single staple is bound to the first domain, the second domain is empty;
	\item	01: a single staple is bound to the second domain, the first domain is empty;
	\item	11: a single staple is bound to both domains;
	\item	12: a distinct staple is bound to each domain.
\end{itemize}
The state of the scaffold is given by the bonding configuration of the domains. Denote $\state = ( p_{0} ,  p_{1}, \ldots , p_{k-1} )$ and let the set of states be $\stateSpace$.
Then the size of the state space is $\left| \stateSpace \right| = 2^{j} \times 5^{l}$ where $j$ is the number of single-domain staples and $l$ the number of two-domain staples.
\teoedit{This description can be extended to staples with more than two domains, although we will focus on the simple case here.   Typically,  each scaffold domain has a unique base sequence, and this restriction is implicit in the state space presented here. In Ref. \cite{Dunn2015}, we show how the state space of the model can be extended to handle scaffolds with repeated sections that allow multiple bonding configurations for each staple.}

\subsection{Kinetic descriptions}
\label{kinetic model}

Simple kinetic models for bimolecular reactions of complementary DNA strands are well established in the literature \cite{Morrison1993, Gao2006}. 
Let complementary strands $A,B$ bind reversibly, forming the double-helical complex $AB$.
Under the assumption of mass-action kinetics, the concentration $[AB]$ is described by 
\begin{align}
	\frac{  d[AB] }{ dt } = k_{+}[A][B] - k_{-} [AB] 
\end{align}
for rate constants $k_{+},k_{-}$. The equilibrium concentrations $\{A\},\{B\}$ and $\{AB\}$ then follow
\begin{align}
	\frac{\{AB\}}{\{A\} \{B\}} = \frac{ k_{+} }{ k_{-} } = \exp \left( \frac{-  \Delta G^{0\,{\rm duplex}}_{AB} }{ R T } \right) \times \mbox{M}^{-1}	\label{ratio} 
\end{align}
where $R$ denotes the molar gas constant and $T$ temperature. 
$	\Delta G^{0\,{\rm duplex}}_{AB} = \Delta H_{AB}^{0\,{\rm duplex}} - T \Delta S_{AB}^{0\,{\rm duplex}}$ is the sequence-specific Gibbs free energy change for duplex formation at molar concentration ($1$\,M),
and $\Delta H_{AB}^{0\,{\rm duplex}}$ and $\Delta S_{AB}^{0\,{\rm duplex}}$ are the corresponding changes in enthalpy and entropy. Throughout this work, superscript $0$ will refer to quantities defined with respect to this standard concentration (the factor of ${\rm M}^{-1}$ arises from dimensional considerations). For a given duplex, these quantities can be estimated through the widely-used
nearest-neighbour model of SantaLucia \cite{SantaLucia1998} in which $\Delta H_{AB}^{0\,{\rm duplex}}$ and $\Delta S_{AB}^{0\,{\rm duplex}}$ are assumed to be $T$-independent.
  
To use a similar description for origami staple binding, we need to establish an expression for $\Delta G^0_{s,s'}$, the difference between the free energies of partially-folded intermediate states $s'$ and  $s$. 
A first approximation to $\Delta G^0_{s,s'}$ might be $\Delta G^{0\,{\rm duplex}}_{s,s'}$, which takes into account only the free energies of hybridized domains calculated as if each hybridized duplex were formed in isolation. However, this ignores significant interactions between different sections of the origami which depend on the state of the folding.
Instead, we take
\begin{align}
	\Delta G^0_{s,s'} = \Delta G_{s,s'}^{0\,{\rm duplex}} + \Delta G_{s,s'}^{\mbox{\scriptsize shape}} + \Delta G_{s,s'}^{\mbox{\scriptsize stack}}, 
	\label{dG_total}
\end{align}
where $\Delta G_{s,s'}^{\mbox{\scriptsize shape}} $ represents the contributions from the entropic costs of scaffold loop formation \cite{Jacobson1950, SantaLucia2004, Arbona2013},
and $\Delta G_{s,s'}^{\mbox{\scriptsize stack}} $ is a contribution from the coaxial stacking of duplex sections \cite{Peyret2000, Pyshnyi2002, Lane1997,Vasiliskov2001}.
We discuss our approximations for the various contributions to $    \Delta G^0_{s,s'}$ in Section \ref{Free energy model}.

The value of $\Delta G^0_{s,s'}$ associated with hybridization of a domain determines the ratio between binding and unbinding transition rate constants, but does not determine their absolute values.
Given $\Delta G^0_{s,s'}$, we calculate the transition rate for the binding of the first arm of a staple as follows. 
Consider an isolated origami in a partially folded state $s_{00}$ with $p_i = 00$ for a particular staple type $i$, and let a staple $i$ bind to the scaffold by a single domain,  resulting in state $s_{01}$ with $p_i = 01$ (Fig. \ref{scaffold}).
We take this transition rate to be equal to that for isolated duplex formation 
\begin{align}
	\rate(s_{00}, s_{01}) = k_{+} [i] ,	\label{Rs00s01} 
\end{align}
where we use $\rate(s,s')$ to denote the rate with which a scaffold makes a transition from state $s$ to $s'$. 
Since the rate of formation of an isolated duplex is known to be more weakly dependent on duplex stability than the corresponding unbinding rate \cite{Morrison1993}, 
we assume that $k_{+}$ is independent of temperature, domain sequence, and folding state, 
and we fix $k_{+} = 10^{6} \mbox{ M}^{-1} \mbox{s}^{-1}$ as a reasonable first approximation \cite{Zhang2009,Morrison1993,Gao2006}. Throughout this work we will assume that the excess staple concentrations are sufficiently high that free staple concentrations can be taken to be constant. \teoedit{Note that this assumption does not mean that staple concentrations are high in absolute terms, only that they are sufficiently in excess of the very dilute scaffold strands.}
The rate $\rate(s_{01}, s_{00})$ for the reverse reaction is then given by:
\begin{align}
	{\rate(s_{01},s_{00})} = k_{+} \exp \left( \frac{ \Delta G^0_{s_{00},s_{01}}}{ RT } \right) \times \mbox{M}.	\label{Rs01s00} 
\end{align}
The binding and unbinding transitions of the second domain of staple $i$ to form $s_{11}$ with $p_i=11$ are associated with the thermodynamic constraint
\begin{align}
	\frac{   \rate(s_{01}, s_{11}) }{ \rate(s_{11}, s_{01}) }  = \exp \left(  \frac{- \Delta G^0_{s_{01}, s_{11}}}{RT} \right) .
	\label{01_to_11/11_to_01}
\end{align}
To resolve the ambiguity in the absolute values of $\rate(s_{01}, s_{11})$ and $\rate(s_{11}, s_{01})$, we make the assumption that 
the unbinding rate $\rate(s_{11}, s_{01})$  is equal to the temperature- and sequence-dependent unbinding rate for the corresponding isolated duplex (with a correction for coaxial stacking): 
\begin{align}
	\rate(s_{11}, s_{01}) = k_{+} \exp \left( \frac{ \Delta G^{0\,\mbox{\scriptsize duplex }}_{s_{01},s_{11}} +  \Delta G_{s_{01},s_{11}}^{\mbox{\scriptsize stack}}  }{ RT } 	\right)  \times \mbox{M}.
	\label{Rs11s01}
\end{align}  
$\Delta G_{ s_{01},s_{11}}^{\mbox{\scriptsize shape}}$ represents the free-energy change corresponding to the change in the geometric constraints imposed by staple crossovers on the part-folded origami. In our model, these constraints are manifest as  
a modified  binding rate for the second staple domain.  Using Eqs. \ref{01_to_11/11_to_01} and \ref{Rs11s01}, we find
\begin{equation}
		\rate(s_{01},s_{11})                                                                                                   
		= k_{+} \exp \left( \frac{ - \Delta G^{\mbox{\scriptsize shape}} _{ s_{01},s_{11}}  }{RT} \right) \times \mbox{M}. 
	\label{internalrate}	
\end{equation}
\teoedit{Eqs. \ref{Rs00s01}, \ref{Rs01s00}, \ref{Rs11s01} and \ref{internalrate} imply that  unbinding transition rates grow rapidly with temperature, whereas binding rates are essentially constant. This feature of the model is consistent with a physical picture in which the barrier to duplex opening is primarily enthalpic, whereas the barrier to duplex formation is primarily entropic.}
  
\subsection{Free energy model} 
\label{Free energy model}
In this section we outline the calculation of various contributions to $\Delta G^0_{s,s'} $. 
\subsubsection{Duplex free energies}
$\Delta G_{s,s'}^{0\,{\rm duplex}}$ is calculated using the well-established SantaLucia parameterization of the nearest-neighbour model of DNA thermodynamics \cite{SantaLucia1998,SantaLucia2004}.
We assume buffer conditions of  $40 \mbox{ mM}$ Tris and $12.5 \mbox{ mM}$ $\mbox{Mg}^{{2+}}$ and apply an additional buffer-dependent entropic penalty for duplex formation \cite{SantaLucia1998, Peyret2000, DNA-Owczarzy-2008-0235}: 
\begin{align}
	\Delta S^{0, {\rm salt}} = 0.368 \times \frac{N}{2} \times \ln \left( \frac{1}{2} [\mbox{Tris}] + 3.3 [\mbox{Mg}^{{2+}}]^{1/2} \right), 
\end{align}
where $N$ is the number of phosphates in the duplex \teoedit{(for a duplex of length $d$ base pairs, we take $N = 2(d-1)$)}. 
  
\subsubsection{Coaxial stacking free energies}
  
It is well known that coaxial stacking of bases across a nick in the DNA backbone can stabilize the flanking duplexes  \cite{Peyret2000, Pyshnyi2002, Lane1997,Vasiliskov2001}.
We add a coaxial stacking contribution $\Delta G^{\mbox{\scriptsize stack}}(T)$ to the free energy of each state wherever two adjacent scaffold domains are both hybridized to staples. The change in coaxial stacking free energy for a transition between states is $\Delta G_{s,s^\prime}^{\mbox{\scriptsize stack}}$.
Geometrical limitations that prevent coaxial stacking, such as bends in the scaffold routing, are ignored.
We parameterize the coaxial stacking strength by $\Delta G^{\mbox{\scriptsize stack}}(T) = n \langle\Delta G^{\mbox{\scriptsize bp}}(T) \rangle $, where $\langle\Delta G^{\mbox{\scriptsize bp}}(T) \rangle $ is the sequence-averaged free-energy gain per base pair in the nearest-neighbour model of Ref. \cite{SantaLucia2004}. \teoedit{$\Delta G^{\mbox{\scriptsize stack}}(T)$, which has been considered by SantaLucia and others \cite{Peyret2000, Pyshnyi2002, Lane1997,Vasiliskov2001}, is much more weakly constrained than other aspects  of the nearest-neighbour model. We thus} treat $n$  as an adjustable parameter of the model, and explore the consequences of varying $n$ in Section \ref{results}. 
  
\subsubsection{Scaffold shape free energies}
\label{shape section}

The presence of $m$ fully-bound, two-domain staples puts $m$ additional constraints on the conformation of the origami, 
\teoedit{as each crossover between staple domains brings non-adjacent scaffold domains into close proximity, pinning
part of the scaffold into a loop.}
The existence of $m$ looping constraints due to $m$ staples suggests that we can decompose $\Delta G^{\mbox{\scriptsize shape}}_{s,s^\prime}$:
\begin{equation}
	G^{\mbox{\scriptsize shape}}_{s} - G^{\rm shape}_{\rm{null}} =  \sum_{L(\state)}{\Delta G_{j}^{\mbox{\scriptsize loop}}} 
	\label{Gshape?}
\end{equation}
where  $L(\state)$ is a set of $m$ loops present in the partially folded structure $s$, and $ G^{\rm shape}_{\rm{null}}$ is the reference conformational free energy for a scaffold with no staples.    
Eq. \ref{Gshape?} is deceptively simple, but before we propose functional forms for $\Delta G_{j}^{\mbox{\scriptsize loop}}$ we must decide which loops should be included in the set $L(\state)$. 
To this end we represent the partially folded origami as a graph in which the circular scaffold is a chain of linked vertices, one for each domain, and each fully-bound staple is an edge connecting the corresponding vertices. We can  identify loops on this graph by finding cyclic paths that do not traverse the same edge twice. The total number of such loops (or simple cycles) that can be drawn, however, grows exponentially with the number of bound staples. It is not obvious how one might algorithmically choose the set of $m$ loops that gives the most physically meaningful representation of $G^{\mbox{\scriptsize shape}}_{s} $ for state $s$. 

\pictureWithCaptionSingleColumn{}{0.99}{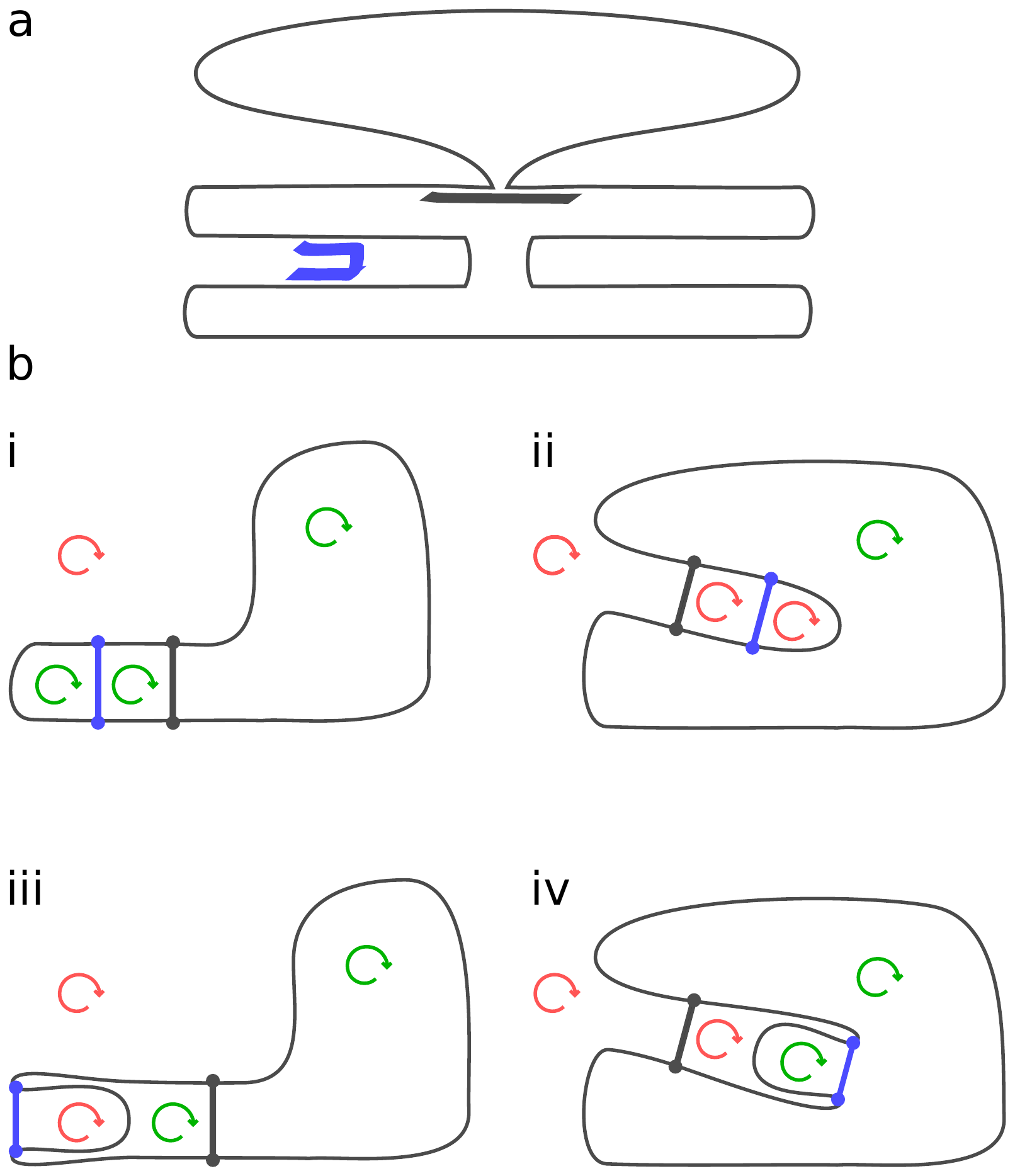}{
	(a): A scaffold with two fully-bound staples.
	(b): Ambiguity in selecting a minimal set of loops from an embedded graph constructed from (a). To construct an abstract graph based on (a), we draw edges that correspond to the scaffold strand and the crossovers between staple domains. Planar projections of this graph ((i)--(iv)) allow the identification of faces that are either internal (green arrows) or external (red arrows) with respect to the scaffold. A staple-free scaffold has two faces (one internal and one external); the addition of each two-domain staple creates an additional face. However, as demonstrated by (i)--(iv), multiple planar embeddings of the same graph are possible, each giving a set of faces that are encircled by distinct loops of DNA. This ambiguity can be resolved by specifying {\it a priori} which staples are external to the scaffold; if black staples are internal and blue external, (b.iii) is the appropriate planar projection.
}

This combinatorial difficulty seems hard to understand when one looks at a schematic representation of a folded origami, such as  Fig. \ref{layout}. It seems almost trivial to  identify a unique set of physically meaningful loops, one for each two-domain staple. 
In the case of flat origami \cite{DNA-Rothemund-2006-0002}, the associated graph has a natural planar embedding given by the two-dimensional shape where the self-evident loops are equal to the faces in the embedding.
Inspired by this observation, we extend the concept of planar embedding to partially-folded states (for a discussion of multi-domain staples, see Appendix \ref{multidomainstaples}). For a given planar embedding, it is easy to identify faces (and hence loops) within an origami, as illustrated in Fig. \ref{situations_fix}. However,  Fig. \ref{situations_fix} also shows that a given graph has multiple possible planar embeddings. To resolve this ambiguity, 
we specify whether the edge associated with each staple lies on the inside or on the outside of the scaffold   
on the basis of the intended origami structure. In  the simple origami  design shown in Fig. \ref{layout}, the staples on the outside of the scaffold are blue, while the inner staples are black. This 
makes the planar embedding of the graph unambiguous at any stage of folding (Fig. \ref{situations_fix}).  We can therefore associate the set of loops $L(\state)$ appearing in Eq. \ref{Gshape?} with the set of loops that enclose the faces of the planar graph of $s$. The addition of any two-domain staple to any partially-formed origami state increases the number of faces in the planar graph by one, and this face can be unambiguously identified.  In Section \ref{loop free energy} we propose a functional form for $\Delta G^{\mbox{\scriptsize loop}}$ which depends on the structure of the DNA loop encircling the relevant face. 
Because a graph with $m$ two-domain staples has $m+2$ faces, we necessarily observe one looping constraint (face) too many, a subtlety we discuss in Appendix \ref{number of faces}. \teoedit{We note that this formalism is not directly related to the terminology of `interior' and `external' loops used in the secondary-structure prediction community \cite{Zuker2003}. Each additional loop in our model, be it assigned to the inside or the outside of the scaffold embedding, represents a physical looping constraint, and would be described as an interior loop in Ref.\,\cite{Zuker2003}.}

In our `global' approach, we recalculate $\Delta G^{\rm shape}_{s,s^\prime}$  exactly (within the approximations of the model) at each stage.
We describe this approach as `global' because we explicitly consider the consequences of an incoming (or outgoing) staple for {\it all} loops, not just the one most obviously associated with the staple in question. If we do not, the resultant dynamics would not reach an equilibrium state characterized by $\Delta G^0_{s,s^\prime}$.  As a simpler alternative, we propose a `local' model for estimating $\Delta G^{\rm shape}_{s,s^\prime}$ in 
which we  include only
the cost of the smallest (minimal) loop that forms or is disrupted during a transition and neglect the effects of the transition on other loops in the graph. In the local model
\begin{align}	
	\Delta G^{\mbox{\scriptsize shape}}_{s,s^\prime}	=	\begin{dcases}
	{\Delta G_{\rm min}^{\mbox{\scriptsize loop}}}  & \mbox{ if a loop forms under } s \rightarrow s^\prime,  \\
	-{\Delta G_{\rm min}^{\mbox{\scriptsize loop}}} & \mbox{ if a loop breaks under } s \rightarrow s^\prime, \\
	0                                               & \mbox{ otherwise. }                                     
	\end{dcases}	     \label{dGshape}
\end{align}
A loop forms (breaks) if a single domain of a two-domain staple binds (unbinds) while the other domain is attached to the origami. We identify the minimal loop as the cycle on the origami graph that has the smallest value of $\Delta G^{\rm loop}$ of those containing the edge that corresponds to the staple crossover in question. To perform this calculation, we use the graph of the state in which the domain in question is bound (for maximal consistency with the global model).

\teoedit{\subsubsection{Illustrating the global and local approaches} 
\label{global v local illustration}
For the sake of clarity, we illustrate which loops are identified when a single staple binds to a simple scaffold (see Fig.\,\ref{example1}), for both the local and global models. Actual reaction rates depend on the parameterization discussed in Section \ref{loop free energy}: example calculations are given in Appendix \ref{computing rates}.

For the global model, changes to each face of the planar graph are evaluated at each step. There is a small but non-zero cost 
\begin{align}
\Delta G^{\rm shape} =  \Delta G^{\rm loop}_{\alpha^\prime} + \Delta G^{\rm loop}_{\gamma^\prime} - (\Delta G^{\rm loop}_{\alpha} + \Delta G^{\rm loop}_{\gamma}) 
\label{dloop}
\end{align}
 associated with the transition from the state in Fig.\,\ref{example1}\,a to  Fig.\,\ref{example1}\,b,  in which the binding of the first arm of a staple alters the physical properties of existing loops, but does not create a new loop. Note that $\Delta G^{\rm loop}_\beta$ does not appear in Eq.\,\ref{dloop} since that loop is unchanged. This  $\Delta G^{\rm shape}$ is  manifest in a changed off-rate (Eq.\,\ref{Rs01s00}). For the transition from  Fig.\,\ref{example1}\,b to  Fig.\,\ref{example1}\,c, the graph topology changes and  
\begin{align}
\Delta G^{\rm shape} =  &\Delta G^{\rm loop}_{\alpha^{\prime \prime}} + \Delta G^{\rm loop}_{\delta} +\Delta G^{\rm loop}_{\epsilon} \nonumber \\
& - (\Delta G^{\rm loop}_{\alpha^\prime} + \Delta G^{\rm loop}_{\gamma^\prime}).
\end{align}
 This  $\Delta G^{\rm shape}$ is manifest in a changed on-rate for the second arm (Eq.\,\ref{internalrate}).

For the local model, $\Delta G^{\rm shape}$  is zero for the transition between the states shown in Fig.\,\ref{example1}\,a and Fig.\,\ref{example1}\,b, and is equal to $\min (\Delta G^{\rm loop}_{\delta}, \Delta G^{\rm loop}_{\epsilon})$  for the binding of the second arm (Fig.\,\ref{example1}\,b to  Fig.\,\ref{example1}\,c), when a new loop is formed. 

Although in principle the two models seem quite different, in practice the numerical values for $\Delta G^{\rm shape}$ are quite similar, which we explain as follows. Firstly, the change in $\Delta G^{\rm loop}$ due to a constituent domain becoming double stranded is generally not that large (see Section \ref{loop free energy} and Appendix \ref{computing rates}). Secondly, as will be discussed in Section \ref{loop free energy}, $\Delta G^{\rm loop}$ is logarithmic in loop length. Thus when a loop is split into two (as in Fig.\,\ref{example1}\,b to  Fig.\,\ref{example1}\,c), creating a larger and a smaller daughter loop, the contribution to $\Delta G^{\rm shape}$ from the smaller daughter is much larger than the difference in $\Delta G^{\rm shape}$ between the original loop and the larger daughter. Therefore simply considering the smallest loop that forms is quantitatively reasonable. }

\pictureWithCaptionSingleColumn{}{0.99}{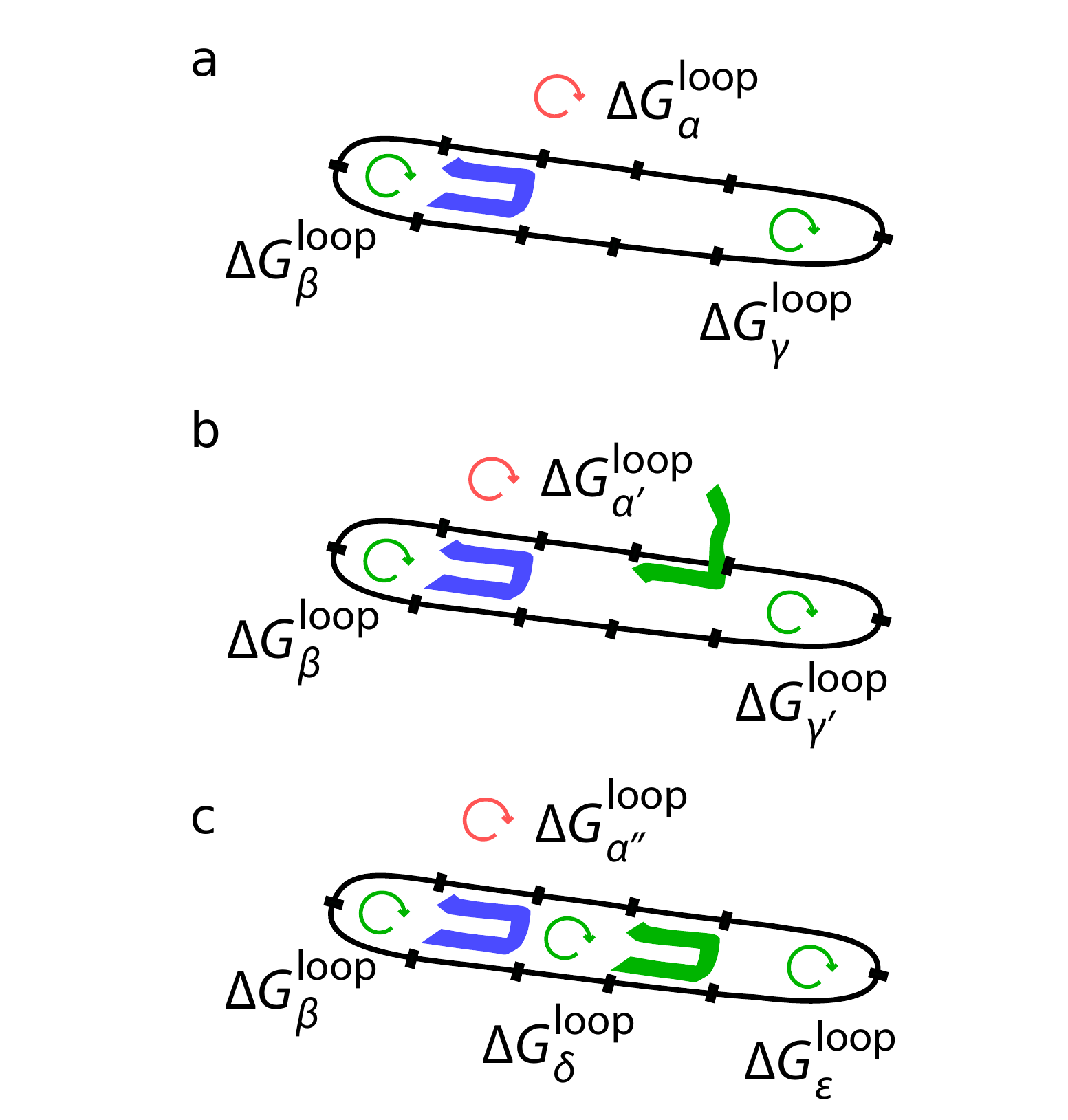}{Loops present in three configurations of a simple origami. 
Loop $\alpha$ comprises the entire scaffold. Additional loops are created by the connections between scaffold domains formed by staples. Primed labels indicate loop costs that have changed as a result of the conversion of a constituent domain from single-stranded to double-stranded DNA.
Topological changes are indicated by changes in the labels
themselves.}

\pictureWithCaptionSingleColumn{}{0.99}{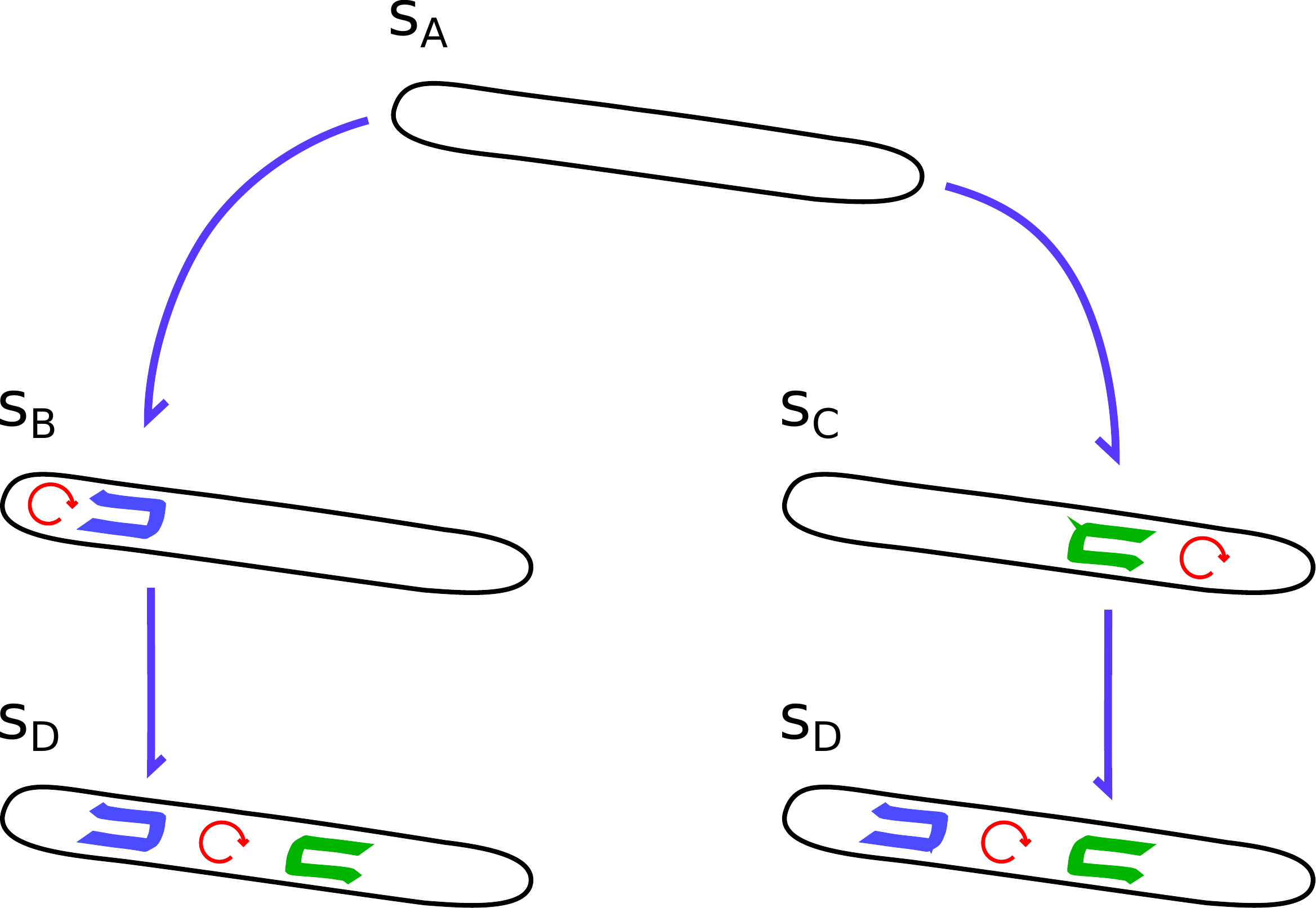}{Thermodynamic inconsistency in the local model: the route $A \rightarrow B \rightarrow D$ produces a different change in free energy from $A \rightarrow C \rightarrow D$. }

\teoedit{
We note, however, that although numerical values of $\Delta G^{\rm shape}$ are similar, there is a fundamental difference between the two approaches. Consider two  staples that can bind to an otherwise empty scaffold (Fig. \ref{inconsistency2}). There are four fully-bound states:
\begin{align}
& s_A = (00, 00), s_B = (00,11),  \nonumber \\
& s_C = (11,00), s_D = (11,11).
\end{align} 
In the global model, each state is assigned a well-defined free energy, including the shape contribution $ \Delta G^{\rm shape}$:  the free energy is
necessarily a function of state and changes in free energy are independent of pathway. However, this is not the case in the local model. Fig. \ref{inconsistency2} indicates free-energy changes associated with
two paths from state A to state D. If the system were rigorously thermodynamically well-defined, as in the global model, then the following equality would necessarily hold:
\begin{equation}
\Delta G^{\rm shape}_{s_A,s_B} + \Delta G^{\rm shape}_{s_B,s_D} = \Delta G^{\rm shape}_{s_A,s_C} + \Delta G^{\rm shape}_{s_C,s_D}
\label{Thermo_equality}
\end{equation}
It is clear that the sets of `minimal' loops used in the local model to calculate the changes in free energy along the two pathways are different, so Eq. \ref{Thermo_equality} will not hold in general in the local model. The local model is therefore not  thermodynamically self-consistent. 
}

Although the local approach does not lead to a well-defined $G^{\rm shape}_s$ for each configuration, 
it is computationally less demanding than the global version 
and has the advantage of supporting non-planar origami designs.  In this work we will compare the two approaches to explore whether the 
local model can reasonably be used to study more complex systems for which the global approach might be impractical. The representation of origami as a graph is discussed further in Section \ref{algorithm}.

\subsubsection{Estimating loop free energy}
\label{loop free energy}
Both the local and global approaches necessitate the calculation of a loop free-energy cost $\Delta G^{\rm loop}$. In this section, we outline a simple estimate of  $\Delta G^{\rm loop}$, and identify the key variables that lead to cooperativity.

Our approach is analogous to that of Jacobson and Stockmayer \cite{Jacobson1950}. We approximate the free-energy cost of loop formation by
\begin{align}
	\Delta G^{\mbox{\scriptsize loop}} = RT \ln   \left(  P^{r_{c}}_{v^{0}} / { P^{r_{c}}_{\mbox{\scriptsize  loop}}}   \right). 
	\label{dGloop}                                                                                                                       
\end{align}
Here, $P^{r_{c}}_{\mbox{\scriptsize  loop}}$ is the probability that the DNA which constitutes a  loop associated with a particular crossover spontaneously adopts a conformation 
in which its ends are within an arbitrary small distance  $r_c$ of each other without being constrained to be there by the crossover. 
$
P^{r_{0}}_{v^0}
$ 
is the probability that two unconnected molecules would be within $r_c$ in a hypothetical ideal system 
of volume $v^0$, where 
\begin{align}
 v^0 	&= \frac{1}{N_{ A}} \times \molar^{-1} 					\\
 N_{A} 	&= 6.022 \cdot 10^{23} \times \mbox{mol}^{-1}.			
\end{align}
In the expression for $\Delta G^{\rm loop}$, $v^0$ arises to correct for the difference between a free energy of association of separate strands under standard molar conditions (which is equal to the hypothetical free energy of association of two isolated strands in a box of volume $v^0$ \cite{Ouldridge_bulk_2010}), and the free energy of forming a loop.  $\Delta G^{\mbox{\scriptsize  loop}}$ thus quantifies the effective concentration of one loop end in the vicinity of the other, 
relative to standard conditions. For a further justification of this approach, see Appendix \ref{DG justification}.

For a loop that is formed in the transition $s_{01} \rightarrow s_{11}$ in the local model, it follows from Eqs. \ref{internalrate}, \ref{dGshape} and \ref{dGloop} that
\begin{align}
	\rate(s_{01},s_{11}) & = k_+ \exp \left(-\Delta G^{\rm shape}_{s_{01},s_{11}}/RT \right)\times {\mbox{ M}},			 		\nonumber                 \\
	                 & = k_+ \exp \left(-\Delta G^{\rm loop}_{\rm min}/RT \right)\times {\mbox{ M}},					 	\nonumber                       \\
	                 & = k_+ \left(\frac{{P^{r_{c}}_{\rm loop}}}{P^{r_{c}}_{v^0}} \right)  \times {\mbox{ M}}, 			\nonumber \\
	                 & = k_+ \left(\frac{{P^{r_{c}}_{\rm loop}}}{4 \pi r_c^3 N_A/3} \right) 
	\label{effective_concentration}
\end{align}
It is clear from comparison with the binding rate of the first arm of a staple that 
$
\exp \left(-\Delta G_{\rm min}^{\rm loop}/RT \right) \times {\rm M}
$ 
is an effective concentration that determines the loop closure rate. Calculation of $\Delta G^{\rm loop}_{\rm min}$ in the local model then reduces to estimation of $P^{r_{c}}_{\rm loop}$, the maximum spontaneous looping probability for a cycle that contains the edge of the staple in question.
    
As a first approximation to $P^{r_{c}}_{\rm loop}$, we treat the DNA that forms the loop as a freely-jointed chain consisting of two distinct segment types, duplexes and single-stranded DNA. 
Let the end-to-end distance of the chain be given by $\distance$, 
denote $P(\distance)$ its probability distribution, 
then $ P^{r_{c}}_{\rm loop}  = \int_{0}^{r_{c}}P(\distance)d\distance$ is the probability that the ends of the loop are within $r_{c}$.
For a chain with $m$ distinct segment types, in the limit of a large number of segments,
\begin{align}\label{PR}
	P(\distance) = 4 \pi \mathsf{\distance}^{2} \left( \frac{3}{2 \pi E[ \distance^{2} ] } \right)^{3/2} 
	\exp \left( \frac{-3{\distance^{2}}}{2 E[ \distance^{2} ] } \right),                         
\end{align} 
where $E[\distance^{2}] = \sum_{i \leq m} N_i b_i^2$ is the mean squared distance between the two ends. Here, $N_i$ is the number of segments of type $i$ with Kuhn length $b_i$. The result for $m=1$ is a classic result of statistical physics \cite{OMFRS1919, Chandrasekhar1943}. The result for $m > 1$ can be understood from the following simple argument. We expect a Gaussian distribution over $X$, $Y$ and $Z$ components of the end-to-end vector for a large number of segments (from the central limit theorem) and the expression for $E[X^2] = E[Y^2] = E[Z^2] = \frac{1}{3}E[\distance^2] $ is trivial to derive for an ideal polymer. We also expect $E[X] = E[Y] = E[Z]=0$, and no correlation between $X$, $Y$ and $Z$ ($E[XY] = E[XZ] =E[YZ]=0$) from symmetry. Only one Gaussian over $X,Y,Z$ satisfies these conditions, and it reduces to Eq. \ref{PR} when expressed in polar coordinates and angular degrees of freedom are integrated over.
   
Substituting the integral of Eq. \ref{PR} into Eq. \ref{dGloop}, we obtain the following expression for $\Delta G^{\mbox{\scriptsize loop}}$ 
\begin{align}
	& \Delta G^{\mbox{\scriptsize loop}}  = RT \ln   \left(  P^{r_{c}}_{v^{0}} / { P^{r_{c}}_{\mbox{\scriptsize  loop}}}   \right) 	 \nonumber                              \\
	                                   & = -RT \ln  \left(	                                                                                                           
	\frac{ \int^{r_{c}}_{0}   {\scriptstyle 4 \pi \distance^{2}} \left( \frac{3}{2 \pi E[ \distance^{2} ]_{\rm loop} } \right)^{\frac{3}{2}}
	\exp \left( \frac{-3{\distance^{2}}}{2 E[ \distance^{2} ]_{\rm loop} } \right)
	d\distance }{ \frac{4}{3} \pi r_{c}^{3} N_{A}  }    
	\right)	\nonumber 	\\	
	                                   & \approx  -RT \ln \left( \frac{ \int^{r_{c}}_{0}   4 \pi \distance^{2} d\distance }{ \frac{4}{3} \pi r_{c}^{3} N_{A}  }                  
	\left( \frac{3}{2 \pi E[ \distance^{2} ]_{\rm loop} } \right)^{3/2}
	\right) 	\nonumber \\
	                                   & = -RT \ln \left( \frac{1}{N_{A}} \left( \frac{3}{2 \pi E[ \distance^{2} ]_{\rm loop}} \right)^{3/2}  \right). 			\label{dGloop_explicit} 
\end{align}  
We have assumed $r_{c}^2 \ll E[ \distance^{2} ]_{\rm loop}$. Thus, in the local model, if a transition involves the binding of the second domain of a staple, and 
$E[ \distance^{2} ]_{\rm min}$ is the minimal value of $E[ \distance^{2} ]$ found for a loop containing that staple,
\begin{align}
	  & \rate(s_{01},s_{11})   \approx \frac{k_{+}}{N_{A}} \left( \frac{3}{2 \pi E[ \distance^{2} ]_{\rm min}} \right)^{3/2}  \label{kifjc} 
\end{align}
Using Eq. \ref{dGloop_explicit} we can write
\begin{align}
	\Delta G^{\mbox{\scriptsize loop}} & = - RT \gamma \ln{ \frac{C}{E[\distance^{2}]_{\rm min}} }		\label{result} 
\end{align}
with $\gamma=3/2$
and 
$
C =   {3}/{2 \pi} \left({1 \times  {\rm M}^{-1}}/{N_{A}}\right)^{2/3}=  6.7 \times 10^{-19} \mbox{ m}^{2}
$.
We express $\Delta G^{\mbox{\scriptsize loop}} $ in this way as it allows us to identify parameters that can be used to generalize our description.
$\gamma =3/2$ is the well-known loop exponent of a freely-jointed chain \cite{Jacobson1950,Fisher1966}. It gives the scaling of the typical volume accessible to the end of a polymer with the polymer's contour length (and  is therefore directly connected to the looping probability). It is well known that  excluded volume interactions tend to swell a polymer chain \cite{Fisher1966}, with the result that an effective $\gamma > 3/2$ is obtained. Theoretical estimates predict $\gamma \sim 1.75$ \cite{Fisher1966} for a self-avoiding walk.  The widely-used SantaLucia model, however, uses a value as high as 
$\gamma = 2.44$ \cite{SantaLucia2004} in an equivalent calculation for the single-stranded bulge loop illustrated in Fig. \ref{bulge}; this estimate is based on a fit to DNA loop closure kinetics \cite{goddard2000}. 
 
$\gamma$ and $C$ play very different roles in $\Delta G^{\rm loop}$. Increasing $\gamma$ at fixed $C$ exaggerates the differences in the loop closure penalty $\Delta G^{\rm loop}$ between longer and shorter loops, whilst also making all loops less stable. Increasing $C$ at fixed $\gamma$ makes all loops more stable by a constant factor.  In this work, we  explore the properties of our model as $\gamma$ and $C$ are modulated. In particular, we consider the consequences of varying $C$ at fixed $\gamma$, and varying $\gamma$ whilst adjusting $C$ so that $\Delta G^{\rm loop}$ for a 18-base single-stranded  bulge loop is fixed at the value obtained in the freely-jointed case.
We couple changes in $\gamma$ and $C$  in this fashion because changing $\gamma$ at fixed $C$ quickly results in unreasonable values of $\Delta G^{\rm loop}$. \teoedit{The choice of a loop length of 18 for calibration is somewhat arbitrary, although it is small enough to be within the range previously tested  \cite{SantaLucia2004}, whilst not being so small that the underlying polymer physics approximations become pathological. The magnitude of $\Delta G^{\rm loop}$ is therefore likely to be physically reasonable at this point.} As an illustration, we plot $\Delta G^{\rm loop}$ for a purely single-stranded loop as a function of length for $\gamma=1.5$, 2.5 and 3.5 in Fig. \ref{fjc2.pdf}.

\pictureWithCaptionSingleColumn{t}{0.7}{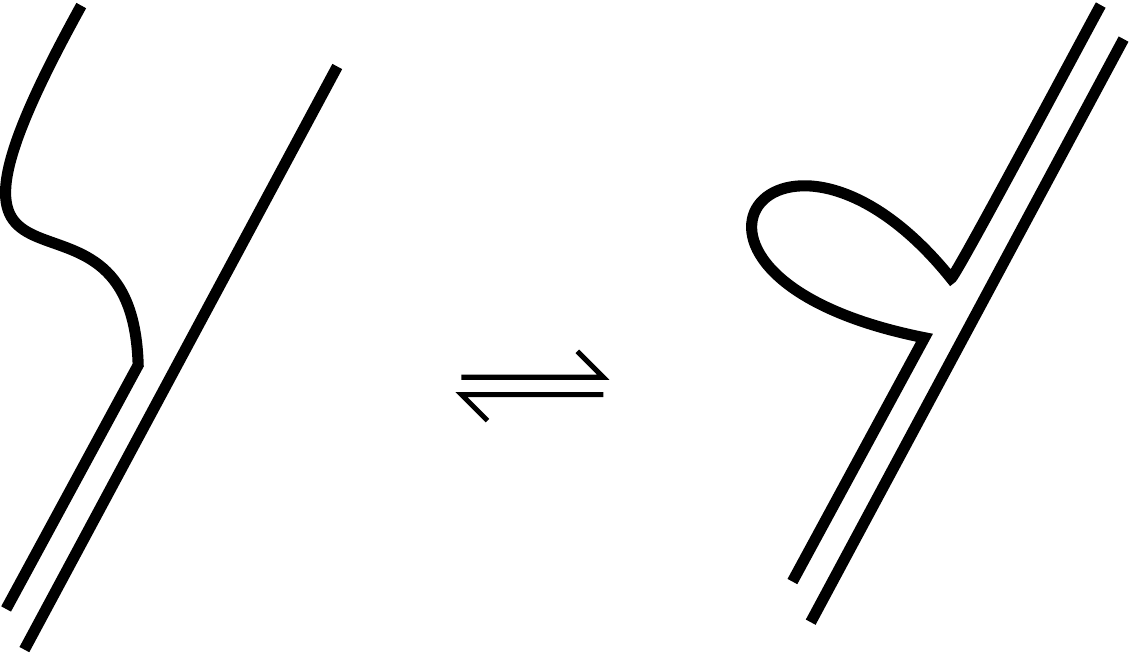}{Bulge formation in a DNA duplex: two long duplex sections enclose a single-stranded loop. Bulge formation is a special case of loop formation within the context of DNA origami. \label{bulge}}

\pictureWithCaptionSingleColumn{t}{0.99}{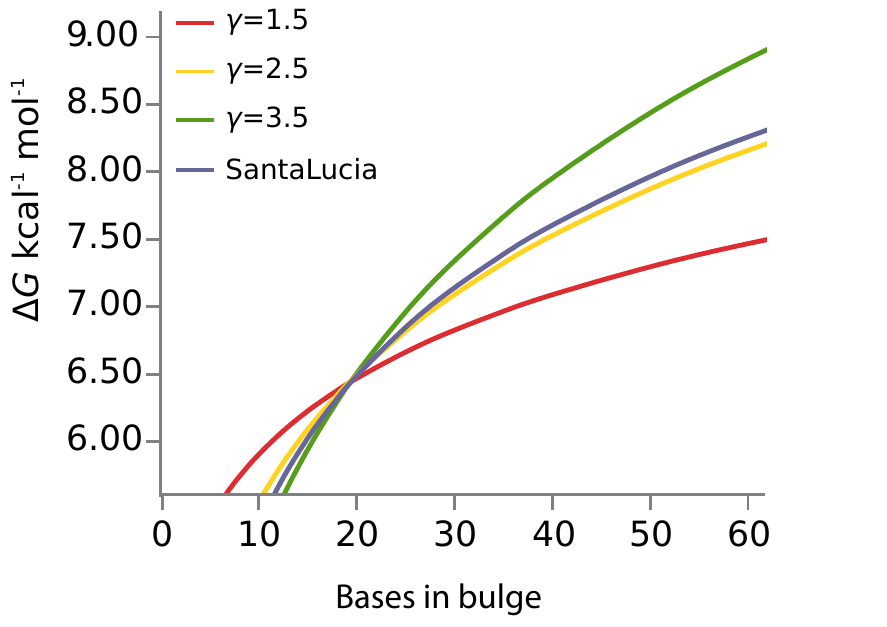}{Comparison between the  free-energy costs of forming a single-stranded bulge loop for different values of the parameter  $\gamma$ at $T=60\,^{\circ}{\rm C}$. A prefactor $C_\gamma$ is adjusted such that $\Delta G ^{\rm loop}$ for an 18-base bulge is held constant. \teoedit{$C_{1.5}=6.7 \times 10^{-19} \mbox{ m}^{2}$,  $C_{2.5}=2.8 \times 10^{-18} \mbox{ m}^{2}$, $C_{3.5}= 5.2 \times 10^{-18} \mbox{ m}^{2}$.} }
    
The mean-squared end-to-end separation of a heterogeneous loop, $E[\distance^{2}]$, is estimated as follows.    
An $x$-base-pair double-stranded domain is treated as a single segment of length $xl_{ds}$, equal to the length of the helix
calculated using a contour length of $l_{ds}=0.34$ nm per base pair  \cite{Saenger}.
We treat an $x$-nucleotide single-stranded domain as consisting of $xl_{ss} / \lambda_{ss}$ segments of length $\lambda_{ss}$,
where $\lambda_{ss}=1.8$ nm is  the Kuhn length of single-stranded DNA, 
and we assume a contour length for the single-stranded section  of $l_{ss}=0.6$ nm per nucleotide. 
These values are roughly consistent with experimental measurements, although the mechanical properties of single strands remain somewhat controversial~\cite{Smith1996,Rivetti1998,   Mills1999, Murphy2004,Chen2012}. 
Where two scaffold domains are held together by a staple, we represent the link by a segment of length $\lambda_{ss}$. \teoedit{The smallest possible loop 
is formed by a ``seam" \cite{Dunn2015}, when two staples connect two pairs of adjacent scaffold domains (for example, the pairs of horizontal black staples in Fig. \ref{layout}). In our model this corresponds to $E[\distance^{2}] = 2 \lambda_{ss}$. This loop cost, combined with duplex initiation terms \cite{SantaLucia2004}, is enough to make the binding of a second seam staple less favourable than a contiguous 32-base-pair domain within our model (at the relevant temperatures for origami assembly), even with coaxial stacking  strength $n=2$.}
 
We emphasize that the model for $\Delta G ^{\rm loop}$ used here is extremely basic and should not be over-interpreted. 
It captures the physics that a staple whose binding sites are connected by a short, flexible loop is more stable than a staple that closes a longer loop.
The model contains physically meaningful parameters that can be adjusted to modulate this effect.  
It allows us to establish a  general framework for modelling that could incorporate optimized estimates of $\Delta G ^{\rm loop}$ in the future. It is thus an excellent tool for our purposes in this article. 

\teoedit{In Ref.\,\cite{Dunn2015}, we used a simplified version of the local model, ignoring sequence-dependent hybridization energies and coaxial stacking, and taking $\gamma=1.5$. Despite these simplifications, the approach taken in Ref.\,\cite{Dunn2015} retains the following key features that underlie the central results: staple insertion rates depend on previously-inserted staples that bring staple binding domains into closer proximity, so staple binding is cooperative; staples that span large scaffold loops are less stable than equivalent short-range staples; and strong cooperative interactions between pairs of such staples provide a compensating stabilization of these long-range connections.}

\subsection{Simulation methods} 
\subsubsection{Algorithm outline}
  \label{algorithm}

The state space $S$, rate matrix $\rate(T(t))$ and initial state $s(t=0)$ form an inhomogeneous continuous-time Markov chain (CTMC) $(S,\rate(T(t)),s(t=0))$, where $T(t)$ is the externally imposed function relating time and temperature that cycles between the initial temperature $T_{\rm{start}}$, the final temperature $T_{\rm{end}}$ and then back to the initial temperature with a fixed rate $|{\rm d}T/dt|$.
Instead of simulating the inhomogeneous CTMC directly, we approximate $T(t)$ and hence the rate matrix $\rate(T(t))$ as piecewise constant across $1$ second intervals, which is a reasonable approximation for typical experimental cooling rates. Individual traces are then generated by applying the standard Gillespie simulation algorithm \cite{Gillespie1976} at each interval.     
To implement the Gillespie algorithm, it is necessary to calculate all transition rates $\rate ( \state,\state^{\prime})$ from the current state $\state$ to alternative states $\state^\prime$.
$\rate(\state,\state^{\prime})$ will be non-zero in the following cases: 

\begin{itemize}
	\item[1] All unbound domains in $s$ can  hybridize with complementary domains of staples present in solution, with a rate given by Eq. \ref{Rs00s01}.
	\item[2] Domains of half-bound staples can unbind, with a rate given by Eq. \ref{Rs01s00}.
	\item[3] Domains of fully-bound staples can unbind, with a rate given by Eq. \ref{Rs11s01}.
	\item[4] Half-bound staples can become fully-bound if the opposing domain is free, with a rate given by Eq. \ref{internalrate}.
\end{itemize}
The term $\Delta G^{0, {\rm duplex}}$, required in transition types 2 and 3, is computed straightforwardly using the nearest-neighbour model of SantaLucia {\it et al.} \cite{SantaLucia2004}.
The term $\Delta G^{0, {\rm stack}}$ is also needed in transition types 2 and 3 and is computed using a simple lookup on the status of neighbouring domains.
The term $\Delta G^{0, {\rm shape}}$ occurs in transition types 2 and 4, and this is where subtleties arise and the global and local models differ. 
In both models, graphs representing states $s$ and $s^\prime$ are needed to calculate $\Delta G^{0, {\rm shape}}$. As the simulation transitions from state to state, the graph is updated. 
In the global model, a specific planar embedding of the graph representing state $s$, as depicted in Fig. \ref{situations_fix}, is used.
This specific embedding is not required in the local model.

The graph $\graph(\state) = (\mathsf{V},\mathsf{E}(\state))$ itself is defined as follows:
each junction between domains on the scaffold is a vertex $v \in \mathsf{V}$ and each domain is an edge $e \in \mathsf{E}(\state)$ between the appropriate vertices. 
Fully-bound staples present in state $\state$ define additional edges between the two vertices that are linked by the staple crossovers. 
A labelling function 
  \begin{align}
    \mathsf{L}:\mathsf{E}(\state) \rightarrow \{ \mbox{single stranded}, \mbox{double stranded},\mbox{crossover} \}
  \end{align}
 assigns the status of each edge, which also has a fixed length (number of nucleotides or base pairs) if it is a scaffold domain rather than a crossover.
 Each edge $e\in \mathsf{E}(\state)$ is weighted as follows:
\begin{align}
\weight(e) =
  \begin{dcases}
    (xl_{ds})^{2} 					& \mbox{if } \mathsf{L}(e) = \mbox{double stranded of length $x$},	\\
      xl_{ss} \lambda_{ss}		 		& \mbox{if } \mathsf{L}(e) = \mbox{single stranded of length $x$},	\\
    \lambda_{ss}^{2} 					& \mbox{if } \mathsf{L}(e) = \mbox{crossover}.	\\
  \end{dcases}
  \label{W(e)}
\end{align}
  The total weight of any loop (simple cycle) within the graph is then $E[\distance^{2}]$, the key quantity in estimating the loop cost (Eq. \ref{result}). 

In the global model, the graph $\graph(\state)$ is assumed to have unique planar embedding given by a set of faces, $\faces(\state)$, that are subgraphs of $\graph(\state)$. That is,
$ \forall \mathsf{F}_{i} \in \faces(\state) : \vertices(\face_{i}) \subseteq \mathsf{V},  \edges(\face_{i}) \subseteq \mathsf{E}(\state) $. 
Each face represents a looping constraint.  The weight of each face is given as 
$ 
  \weight(\face_{i}) = \sum_{\edge \in \edges(\face_{i})} \weight(\edge)
$.
The shape contribution to the free energy is thus
\begin{align}
  G^{\mbox{\scriptsize shape}}_{s} - G^{\rm shape}_{\rm{null}} = {   -RT\gamma \sum_{ \face_{i} \in  \faces(\state) }{  \ln \frac{C}{\weight(\face_{i})}}    }    
\end{align}
where the set of loops in \eqn \ref{Gshape?} is substituted with the faces in the embedding of $\graph(\state)$.
As the simulation progresses, the faces of the graph are merged (transition type~3) or split (transition type~4),
and we use a custom data structure to dynamically update the faces of the graph.
During a transition, 
it is only necessary to recalculate $E[\distance^{2}]_{\rm loop} = \weight(\face_{i})$  and  $\Delta G^{\rm{loop}}$ for the affected loops (faces), which are easy to identify.

The local model does not use an embedded graph representation but does use the same weighted graph $\graph ({\state})$. 
Estimating the change of $\Delta G^{0,\rm{shape}}$ is only necessary for full binding of a previously half-bound staple (transition type~4): in that case we approximate 
$  \Delta G^{0,\rm{shape}} = \Delta G^{\rm loop}_{\rm min}$,
where $\Delta G^{\rm loop}_{\rm min}$ is the minimal $ \Delta G^{\rm loop}$ of any loop incorporating the newly formed staple crossover in the new state $s^\prime$.
This corresponds to finding the simple cycle including the new crossover that minimizes $E[\distance^{2}]$.
We employ Dijkstra's shortest path algorithm \cite{Dijkstra} to find the shortest path in $s^\prime$ between the two vertices 
that are to be connected by the staple crossover (excluding the crossover itself). This path is then added to the crossover in question to make the shortest loop. 
Given a state $\state$ and a half-bound staple $p$, let $\vertex_{1},\vertex_{2}$ be the vertices that are joined by a new edge once $p$  becomes fully bound by hybridization to domain $\edge$. 
Let the new graph $\graph^{\prime}$ be equal to $\graph(\state)$ except that $\status(\edge) = \mbox{double stranded}$, and let $\mathsf{D}(\vertex_{1},\vertex_{2})$ be the weight of the shortest path between $\vertex_{1},\vertex_{2}$ in $\graph^{\prime}$ under $\weight$.
Then 
\begin{align}
\Delta G^{\rm{shape}}_{\min} 	&= -
RT\gamma \ln \frac{C}{E[\distance^{2}]_{\rm min}}			\\
  E[\distance^{2}]_{\rm min} 	&= \lambda_{ss}^{2} + \mathsf{D}(\vertex_{1}, \vertex_{2}).
\end{align}
Example loop calculations for both the global and local models are given in Appendix \ref{computing rates}. The simulation code itself is found at \url{https://github.com/fdannenberg/dna}.

\subsubsection{Computational Tractability}
\teoedit{For the origami studied in this paper, using  $\gamma = 2.5$ and $n = 2$ and a temperature gradient of 1.0$\,^\circ$C\,min$^{-1}$,
160 simulated folding and melting trajectories between 80$\,^\circ$C and 20$\,^\circ$C take 21 minutes (local model) or 30 minutes (global model)  with 10 parallel threads on workstation hardware (IntelTMXeon R
X5660). All other factors being equal, simulating a larger structure with more domains is more computationally demanding, and the number of binding or unbinding events in a given unit of physical time should scale linearly with the number of domains. At least for the Gillespie algorithm, a linear increase in the number of transitions gives a linear contribution to the scaling of the simulation time. 

Our software simulates our origami tile at acceptable speed and we have not attempted to further optimize our code, or establish the scaling of the cost per transition with system size.
The frequently-used M13 genome is approximately three times the length of our scaffold; such an increase in size should not render the system intractable. Indeed, we simulated a system of twice the size of that considered here  in Ref. \cite{Dunn2015}, using a variant of the local model.
 We note, however, that other factors may be at least as important as system size in determining the computational challenge. Reaction rates are sensitive to staple concentration and domain hybridization free energies and faster reactions lead to more transitions. In particular, if domains of very different stabilities are part of the same staple, the less stable domain would be expected to bind and unbind many times before reaching a temperature at which it is stable, leading to a stiff simulation. For a  large origami with strongly heterogeneous domains, therefore, it may be necessary to carefully profile and optimize the computational protocol.
Caching of generated transition rates, using a hash-table and an efficient hashing function (where states are keys and lists of transitions are values), was found to have a low hitting rate for our simulation, but may significantly benefit the mentioned stiff models. Upper bounds on the algorithmic complexity of our simulation can be found by considering each graph query in isolation. However, our simulation is dynamic, where graph structures are updated by inserting or removing one edge at a time, and many queries are performed on the same graph: the proposed upper-bound would not be indicative of the actual problem, and proper theoretical treatment lies outside the scope of this work.
}

\pictureWithCaption{!}{0.98}{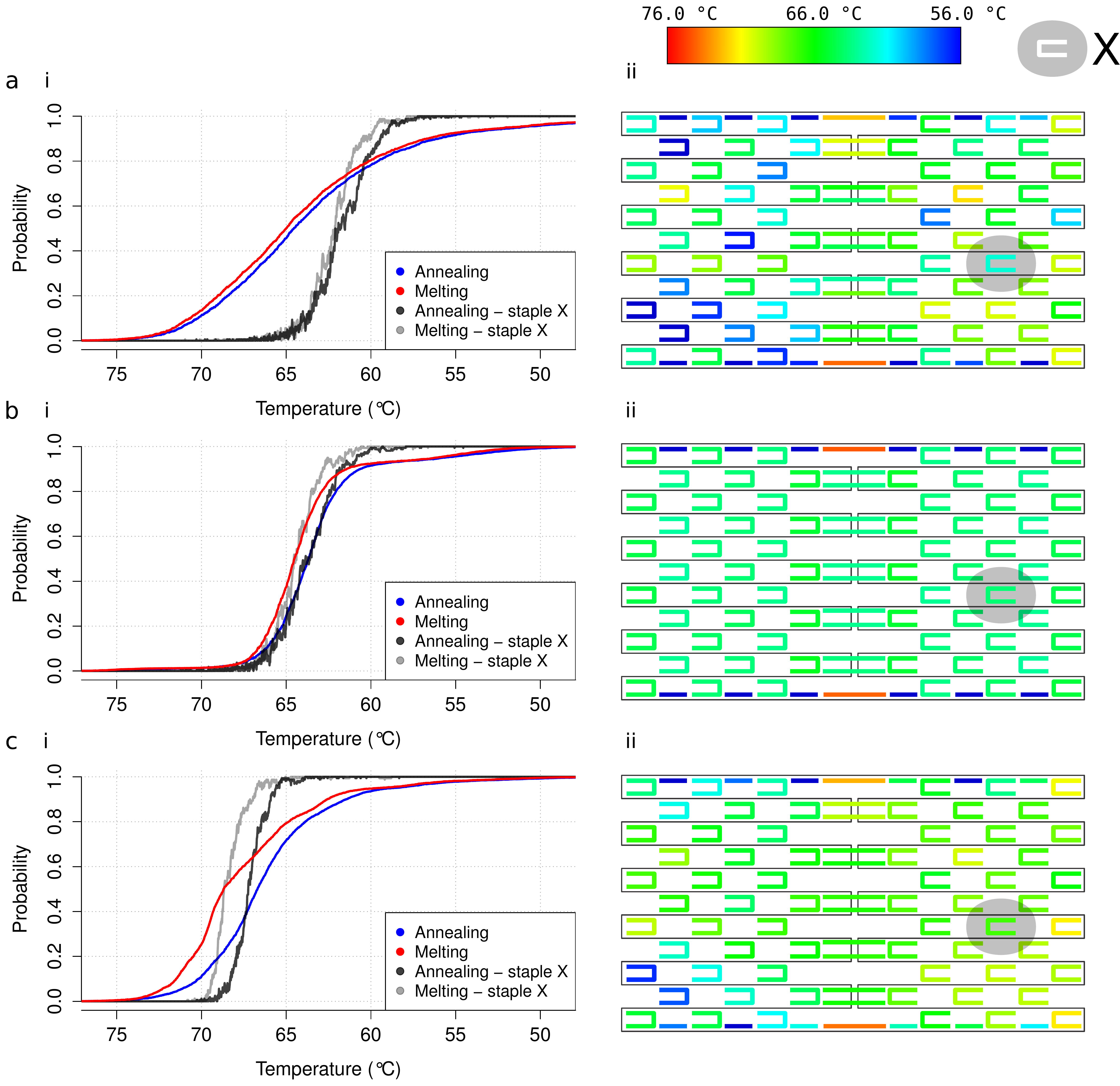}{ 
  Staple binding as a function of temperature during cooling and heating of a simple origami in the global model. In column i we show the average number of scaffold domains that are occupied as a function of temperature. Curves are averages over $160$ cycles with a temperature ramp of $1\,^{\circ}\mathrm{C} $ per min.  Also shown is the probability of two-domain binding for a specific staple $X$ which is highlighted in the schematic shown in column ii.
  In column ii we indicate the incorporation temperature of each staple (the highest temperature for which the probability of full binding is $>50\%$ during annealing). Staples with a low incorporation temperature $<56\degreeC$ are coloured dark blue.  
  (a) Simulation using different (sequence-specific) duplex binding energies for each domain, $\gamma=1.5$ and coaxial stacking strength $n=0$. 
  (b)  As (a), except that all 15/16 base-pair domains are taken to have $\Delta G^{0\,{\rm duplex}}(T)$  equal to the average for a 16 base-pair domain. 
  (c) As (a) but  using $\gamma = 2.5$ and coaxial stacking strength $n=2$. 
\label{Basic_behaviour}
}      

\gnuplotpicturesWithCaptionSingleColumn{!}{0.7}{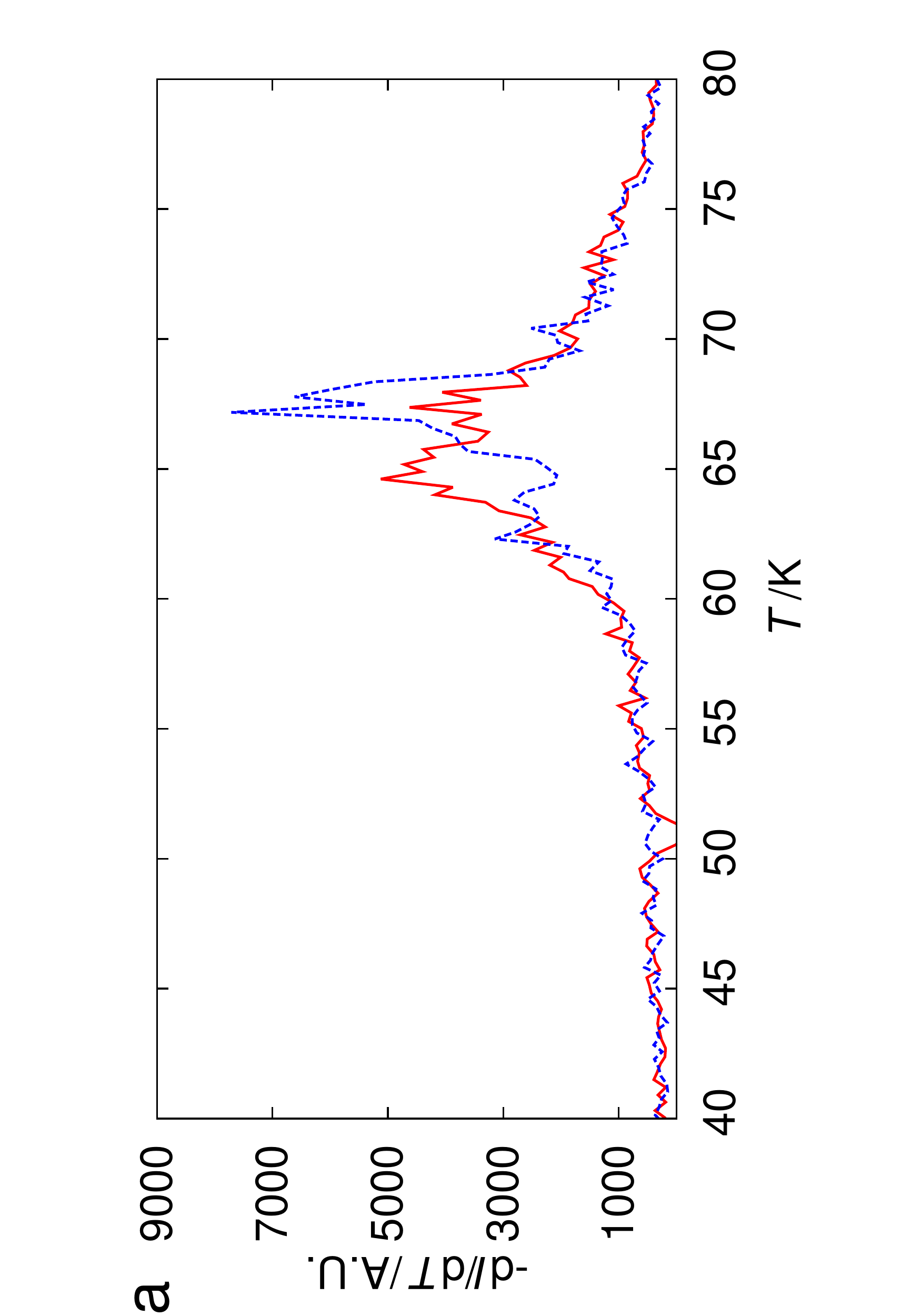}{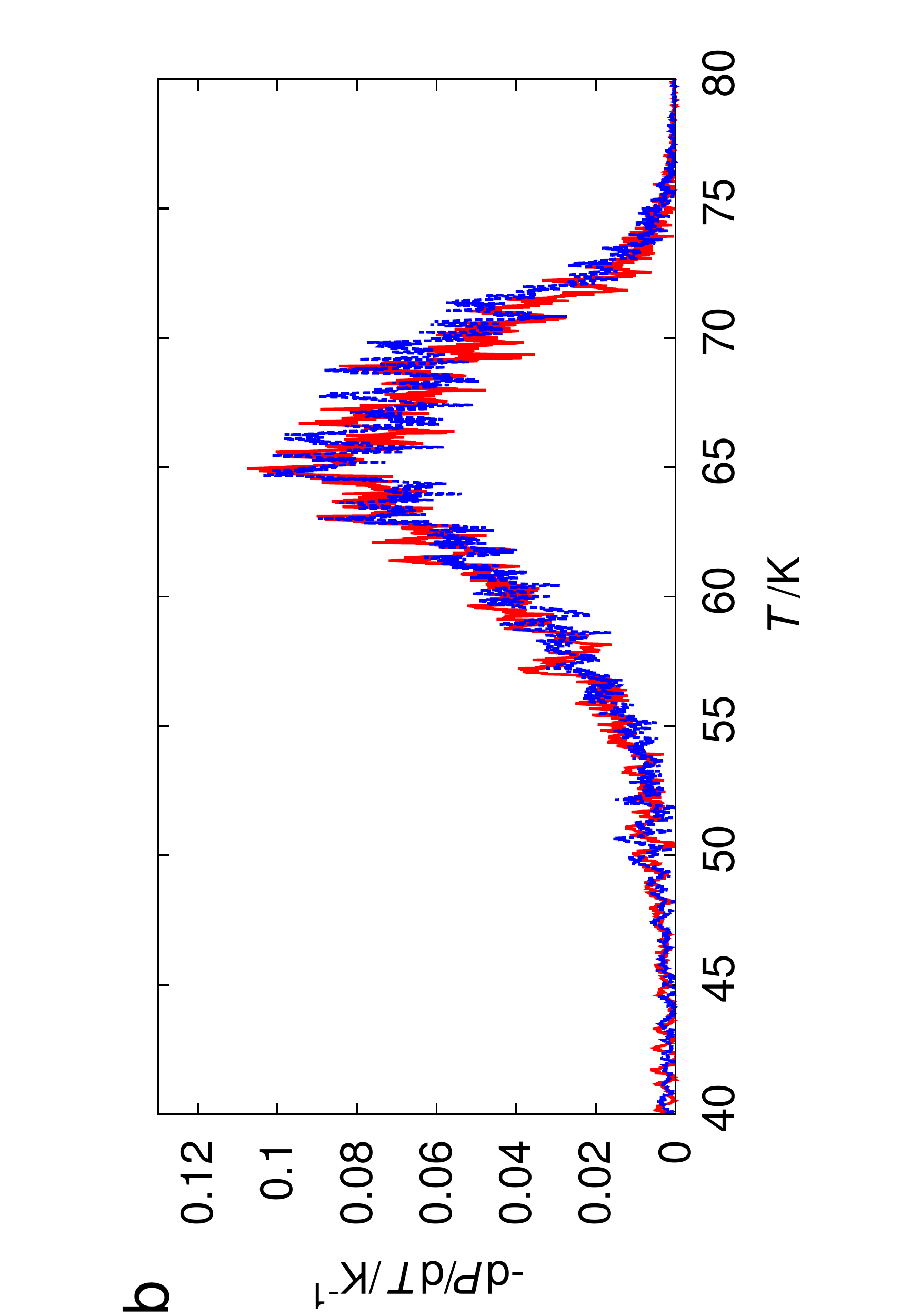}{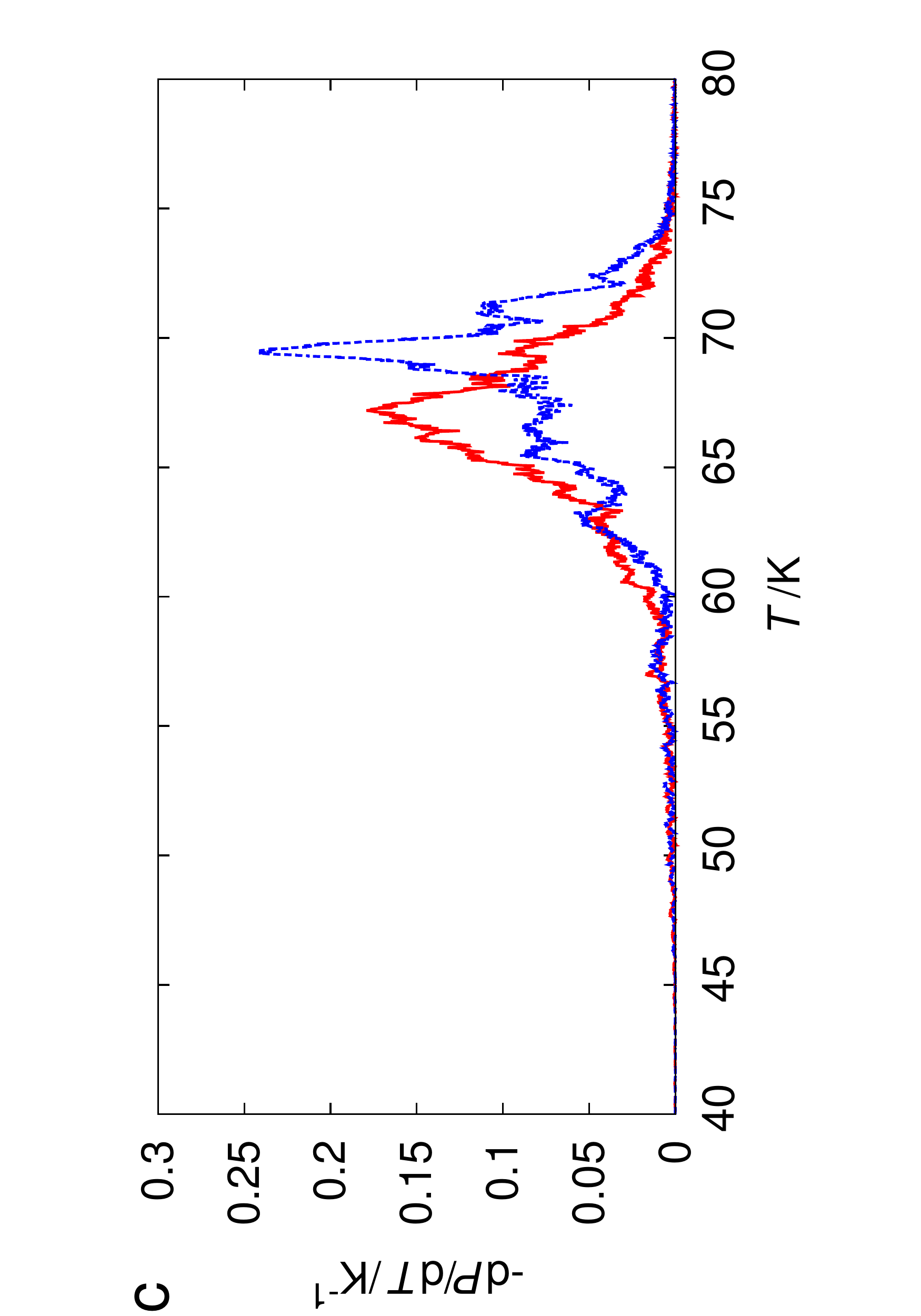}{ 
\teoedit{Comparison of an experimental measurement of  origami folding, as reported by $\sybr$ Green fluorescence $I$, and model behaviour. (a)  $-{\rm d} I/{\rm d} T$ is reported against $T$ during cooling (red, solid) and heating (blue, dashed) between $80\degreeC$ and $40\degreeC$ at a rate of $1\degreeCpMin$, for a single experimental system of $10$ nM scaffold, $20$ nM of each staple and $1\times$ $\sybr$ Green. $I$ is a proxy for the degree of hybridisation. In (b) and (c) we plot $-{\rm d} P/{\rm d} T$, where $P$ is the average number of scaffold domains occupied, obtained with the global model under the same conditions, averaged over 160 realizations. (b) $n=0$, $\gamma=1.5$. (c) $n=2$, $\gamma=2.5$. Plotting ${-\rm d} I/{\rm d} T$ rather than $I$ makes comparison easier without adjusting for baseline effects. All curves are smoothed to improve visibility: experimental data is averaged over a 72\,s window, corresponding to a temperature change of $1.2\degreeC$, and simulation data over a window of $0.3\degreeC$. }}{jb_fig.pdf}

\subsection{Experimental methods}
\label{exp_methods}

To explore whether the model gives a reasonable description of origami assembly, we report experimental measurements of annealing and melting for the same scaffold and staples that we simulate with the model. Origami folding is followed using $\sybr$ Green, a dye whose fluorescence is strongly enhanced when bound to duplex DNA. We monitor fluorescence during repeated cooling and heating cycles of the solution containing scaffold, staples and $\sybr$ Green. 
A scaffold concentration of $10$ nM was used, with each staple present at $20$ nM. During folding, the staple concentration is therefore depleted by a factor of two (the model treats it as constant). 
The likely result is a slightly broader annealing transition during the experiment than in the limit of negligible scaffold concentration.

\section{Results and Discussion}
\label{results}
    
\subsection{ Initial simulations }	
We first display the behaviour of the global model using $\gamma=1.5$ and without coaxial stacking stabilization ($n=0$). To do this, we simulate the folding of a simple origami as illustrated in Fig. \ref{layout}. This origami contains 166 domains, each of 15 or 16 base pairs (apart from two long domains of 32 base pairs); the sequences of scaffold and staples are equal to that of the `monomer tile' in Ref. \cite{Dunn2015}. 
All staples are at a concentration of 20\,nM, and the scaffold strands are assumed to be dilute enough that reduction of staple concentration during folding can be neglected.   We simulate 160 independent folding trajectories using the Monte Carlo algorithm outlined in Section \ref{algorithm}. In each case the system  temperature is reduced by  $1 \degreeCpMin$ from $85 \degreeC$ to $25 \degreeC$, 
at which point the temperature is cycled back to $85 \degreeC$ at the same rate.

Fig.\,\ref{Basic_behaviour}\,(a) shows the average fraction of domains that are bound as a function of temperature during both annealing and melting, together with the degree of incorporation of a single typical staple (staple $X$). Origami assembly occurs at $T_a \approx 65 \degreeC$. This midpoint of the annealing transition is approximately consistent with that observed in the equivalent experimental system (Fig. \ref{jb_fig.pdf}). The model shows very little hysteresis: annealing and melting curves nearly overlap, despite the rapid rate of cooling. Hysteresis is a generic feature of origami systems \cite{Sobczak2012,Wei2013,Arbona2013}, and significant hysteresis is also seen in our experimental measurements (Fig. \ref{jb_fig.pdf}). It is clear that the basic model with $\gamma =1.5$ and without coaxial stacking fails to capture this effect.  

The simulated melting and annealing transitions are fairly broad: the transition from $20\%$ of domains bound to $80\%$ of domains bound during annealing occurs over a temperature range of  $\Delta T_a \approx 9 \degreeC$. \teoedit{Transition widths inferred from experimental data (particularly in the case of melting) are clearly sharper than predicted by the model, as shown in Fig. \ref{jb_fig.pdf}.}  

Interestingly, the widths of the simulated folding and unfolding transitions are not reflected in the corresponding curves for individual staples. For example, staple $X$ goes from 20\% to 80\% bound over a temperature range of  $\Delta T^X_a \approx 2.5\degreeC$. Rather, this width represents the range of incorporation temperatures for individual staples; some staples are more stable than others due to differences in sequence and between the loops that they must enclose. These differences are evident in the heatmap of incorporation temperatures in Fig.\,\ref{Basic_behaviour}\,(a.ii)). In Fig. \ref{Basic_behaviour}\,(b), we consider a system in which all domains (except the two longer domains of 32 base pairs) are assigned the same 
$\Delta G^{0\,{\rm duplex}}(T)$. $\Delta T_a $ is drastically reduced, and is not much larger than $\Delta T^X_a$
if outlying single-domain staples are excluded. The narrow range of incorporation temperatures for all two-domain staples in this case is very clear from Fig.\,\ref{Basic_behaviour}\,(b.ii).
    
\begin{table}[!]
	\setlength{\tabcolsep}{6pt}
	\begin{tabular}{l|rrr|rrr}
		\multicolumn{7}{l}{ \quad \quad  $T_{a}$/$\degreeC$  } 	\\
		\multicolumn{7}{l}{} 										\\	
		Coaxial stacking		   &	\multicolumn{3}{c|}{Global model - $\gamma$}	&	\multicolumn{3}{c}{Local model - $\gamma$ }	\\
		 $n$	   & $1.5$ & $2.5$  & $3.5$ 	& $1.5$ & $2.5$    & $3.5$ \\
		\hline
		0                  & 64.6  & 63.4   & 61.9 	& 64.6  & 63.2     & 61.7  \\
		1.0                & 66.6  & 65.2   & 63.8 	& 66.4  & 65.1     & 63.8  \\
		2.0                & 68.0  & 66.7   & 65.5 	& 67.9  & 66.6     & 65.5  \\
		3.0                & 69.1  & 67.9   & 66.7 	& 69.0  & 67.8     & 66.7  \\
		\multicolumn{7}{l}{} 							\\
		\multicolumn{7}{l}{  \quad \quad  Hysteresis/$\degreeC$ } 	\\
		\multicolumn{1}{l}{} 							\\
		\hline
		0                  & 0.1  & 0.4     & 0.8 	& 0.2   & 0.4     &  0.6 \\
		1.0                & 0.6  & 1.0     & 1.3  	& 0.7   & 0.8     &  1.0  \\
		2.0                & 1.5  & 1.8     & 1.9  	& 1.4   & 1.7     &  1.7 \\
		3.0                & 2.2  & 2.5     & 2.7  	& 2.3   & 2.5     &  2.4 \\
		\multicolumn{7}{l}{} 							\\
		\multicolumn{7}{l}{ \quad \quad $\Delta T_{a}$/$\degreeC$ } 				\\
		\multicolumn{1}{l}{} 							\\
		\hline
		0                  & 8.7   & 8.6   &  8.6  	& 8.8   &  8.8    &  9.1  \\
		1.0                & 6.6   & 6.5   &  6.5  	& 6.7   &  6.7 	  &  6.8  \\
		2.0                & 5.2   & 5.1   &  5.0 	& 5.4   &  5.3    &  5.2  \\
		3.0                & 4.4   & 4.2   &  4.3  	& 4.4   &  4.3    &  4.3  
	\end{tabular}
	\caption{Model properties as functions of exponent $\gamma$ and coaxial stacking strength $n$. 
		Data reported,  averaged for all simulation runs, are  annealing temperature $T_a$ (the first point at which a moving average of at least 50\% of domains are hybridized during annealing), hysteresis ($T_m - T_a$, where $T_m$ is an equivalent quantity to $T_a$ defined during melting), and transition width $\Delta T_a$ (the range of temperatures over which domains go from 20\% bound to 80\% bound during annealing). 
		Data is presented for heating/cooling rates of $1 \degreeCpMin$ for the origami shown in Fig.\,\ref{layout}. Standard errors on the mean, estimated from 160 independent cycles, are smaller than $0.1 \degreeC$ for all data reported.
		\label{table1}	
	}
\end{table}

\subsection{Exploring parameter space}
We now explore the effects of varying model parameters. When specifying the model in Section \ref{the model}, the key quantities that were  left as explicitly variable were loop parameters $\gamma$ and $C$ (see Eq.\,\ref{result}), and the strength of coaxial stacking $\Delta G^{\mbox{\scriptsize stack}}(T) = n \langle\Delta G^{\mbox{\scriptsize bp}}(T) \rangle $. Table \ref{table1} shows the variation in the annealing temperature, hysteresis, and annealing transition width with $\gamma$ and $n$. Note that $C$ is varied with $\gamma$ to ensure a constant cost for a 18-base bulge as discussed in Section \ref{loop free energy}.

The following trends are clear.
\begin{enumerate}
	\item Increased coaxial stacking leads to higher melting and annealing temperatures, sharper transitions and increased hysteresis.
	\item Increased $\gamma$ leads to lower annealing temperatures and increased hysteresis, but has only a weak effect on transition widths.
\end{enumerate}
Fig. \ref{Basic_behaviour}\,(c) shows a combination of these effects for $\gamma=2.5$ and $n=2$. 

The consequences of coaxial stacking are easiest to understand. Coaxial stacking stabilizes the origami structure, and it is therefore unsurprising that it increases $T_a$. It is a cooperative interaction, meaning that the binding of one staple favours the subsequent binding of another. Cooperativity tends to result in narrower transitions, as the binding of isolated staples is suppressed relative to the formation of well-folded regions.
As partially-formed regions are relatively disfavoured, cooperativity tends to exaggerate hysteresis (the system must pass through more substantial free-energy barriers to assemble or melt, slowing down kinetics). 

The influence of $\gamma$ is more subtle. The net effect of increasing $\gamma$ is to increase the  penalty for loop closure, $\Delta G^{\rm loop}$, for longer loops relative to that for shorter loops.  We adjust $C$ to maintain the penalty for a 18-base ssDNA loop: for the origami folding process studied here, the majority of loops are longer than this so the average free energy of loops is increased, explaining the drop in $T_a$ with increasing $\gamma$.  The presence of previously-bound staples can substantially reduce the lengths of the loops closed by incoming staples, and thus the entropic cost of loop closure,  and might thus be expected to lead to stronger cooperative effects.  Consistent with this hypothesis, we see that increased $\gamma$ leads to larger hysteresis. Transition widths, however, show only a weak dependence on $\gamma$. 

Two effects reduce the dependence of transition widths on $\gamma$ by counteracting the expected narrowing as a consequence of enhanced cooperativity. Firstly, increasing $\gamma$ has the effect of increasing the intrinsic differences in stability between staples  by increasing the sensitivity of the loop-closure penalty to loop length. This  tends to make the overall annealing transition {\it broader} if no other effects come into play. Consistent with this, if we assign a state-independent $\Delta G^{\rm loop}$ to each two-domain staple (by calculating the loop cost for a staple binding to an otherwise empty origami, and not updating this value as other staples bind), we observe that  $\Delta T_a$ systematically increases with $\gamma$ (see Appendix \ref{additional data}).

\begin{table}[]	
	\begin{tabular}{l|c|c|c}
		$C_{5/2}'$           & {$ T_{a}$/$\degreeC$} & Hysteresis/$\degreeC$  & {$ \Delta  T_{a}/\degreeC$} \\
		\hline
		$0.5 \times C_{5/2}$ &  65.3	&  1.9		& 5.0		\\
		$1.0 \times C_{5/2}$ &  66.7	&  1.8		& 5.1           \\
		$2.0 \times C_{5/2}$ & 	68.2	&  1.6 		& 5.2           \\
	\end{tabular}
	\caption{ Effect of varying $C_{\gamma}$ on the hysteresis, annealing temperature and width of the transition curve, using $\gamma=5/2$ and coaxial stacking equivalent to 2bp.
		\label{table2}
	}
\end{table}

Secondly, there is a competing anti-cooperative effect, mediated by  loops, that grows with $\gamma$. $\Delta G^{\rm loop}$ grows with the expected end-to-end separation $\surd{E[\distance^{2}]}$ of the DNA that forms the loop. One way to reduce this cost is to create a shortcut through binding a two-domain staple across the loop (the cooperative effect discussed above). Alternative transitions, however, transform a scaffold domain within the loop from single-stranded to double-stranded DNA {\em without} creating a shortcut. Due to the increased stiffness of double-stranded DNA, converting a section of ssDNA to dsDNA always increases  $E[\distance^{2}]$ and hence $\Delta G^{\rm loop}$. Increasing $\gamma$ can thus make some staples less cooperative (or more anti-cooperative), reducing the impact of $\gamma$ on $\Delta T_a$. We note that the model of Arbona {\it et al.} reduces the cost of loop formation by an amount proportional to the number of duplex base pairs within the loop \cite{Arbona2013}, although the reason for this is unclear. 
    
In Table \ref{table2}, we show the results of varying the parameter $C$ (Eq.\,\ref{result}) at fixed $\gamma$ and coaxial stacking strength $n$. The dominant effect is that annealing temperature rises with $C$; comparatively small changes are seen in the degree of hysteresis and the width of the annealing transition. This behaviour is reasonable: from Eq.\,\ref{result} it is clear that $C$ gives rise to a loop-independent contribution to $\Delta G^{\rm loop}$, and hence its dominant effect is to modify the stabilities of all two-domain staples in a systematic fashion.

For the rest of this work, we take $\gamma=2.5$ and $n=2$ as parameter values which match  the scale of hysteresis and cooperative effects seen in the experimental data of Fig.\,\ref{jb_fig.pdf}.
Our value for exponent $\gamma$ is close to that used by SantaLucia \cite{SantaLucia2004}. There is limited data on the stabilizing effects of coaxial stacking, but reported values of $\Delta G^{\rm stack}$ at $T=37 \degreeC$ and $1 \mbox{ M} \mbox{ Na}^+$ are about $15\%$ larger in magnitude than the contribution of one base pair \cite{Pyshnyi2002,Pyshnyi2004}. It is not clear how this parameter changes as the temperature increases towards $65 \degreeC$, the temperature around which  annealing/melting occurs. Our parameters are not fitted or optimized, but rather a reasonable choice that allows us to study generic phenomena in origami folding. \teoedit{More experimental data would allow refinement of our basic description of the contributions to the free energy of a part-folded origami, both in terms of the parameters $\gamma$ and $n$ and more generally with regard to the functional form of our free energy.}

\subsection{Global and local models}
In Section \ref{shape section} we introduced two alternative approaches: the global and local models. Data presented hitherto has been for the global model. The local model is a simpler but less rigorous alternative, which may prove useful in modelling more complex origami structures. To use the local model with confidence, however, it is important to establish how closely it matches the global approach.

Results from both models are presented in Table \ref{table1}. Their predictions are quantitatively  similar, and most physically relevant trends are reproduced. There are systematic differences, however. Most noticeably, the local model predicts a weaker dependence of hysteresis on $\gamma$ (Appendix \ref{additional data}). As discussed in Section \ref{global v local illustration}, the local model cannot capture all staple interactions mediated via loops, and so it is perhaps unsurprising that the most noticeable differences between the two approaches should be manifested in the response to $\gamma$.

\pictureWithCaption{bt}{0.98}{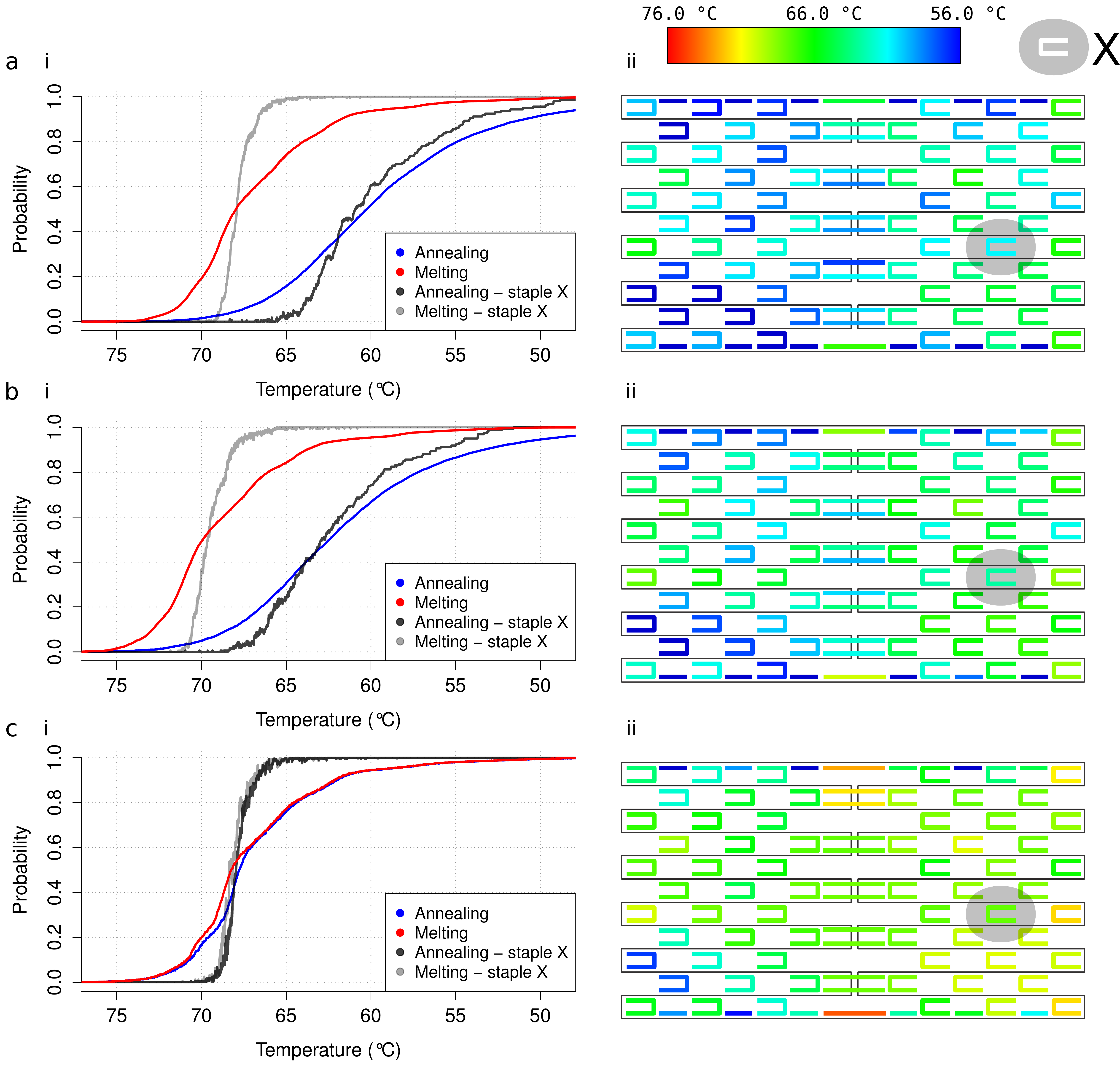}{ 
Simulations of the global model for different staple concentrations and heating/cooling rates. In all cases, we use $\gamma=2.5$  and coaxial stacking strength $n=2$. In column i 
we show the average number of scaffold domains that are occupied as a function of temperature. Curves are averages over $160$ cycles with a temperature ramp of $1\,^{\circ}\mathrm{C} $ per min. 
Also shown is the probability of two-domain binding for a specific staple $X$ highlighted in the schematic shown in column ii.
In column ii we indicate the incorporation temperature of each staple (the highest temperature for which the probability of full binding is $>50\%$ during annealing).
Staples with a low incorporation temperature $<56\degreeC$ are coloured dark blue. 
(a) Staple concentration of $2$ nM, cooling rate of $1\degreeCpMin$. 
(b) Staple concentration of $20$ nM, cooling rate of $10\degreeCpMin$. 
(c) Staple concentration of $20$ nM, cooling rate of $0.1\degreeCpMin$. 
}   

\pictureWithCaption{bt}{0.98}{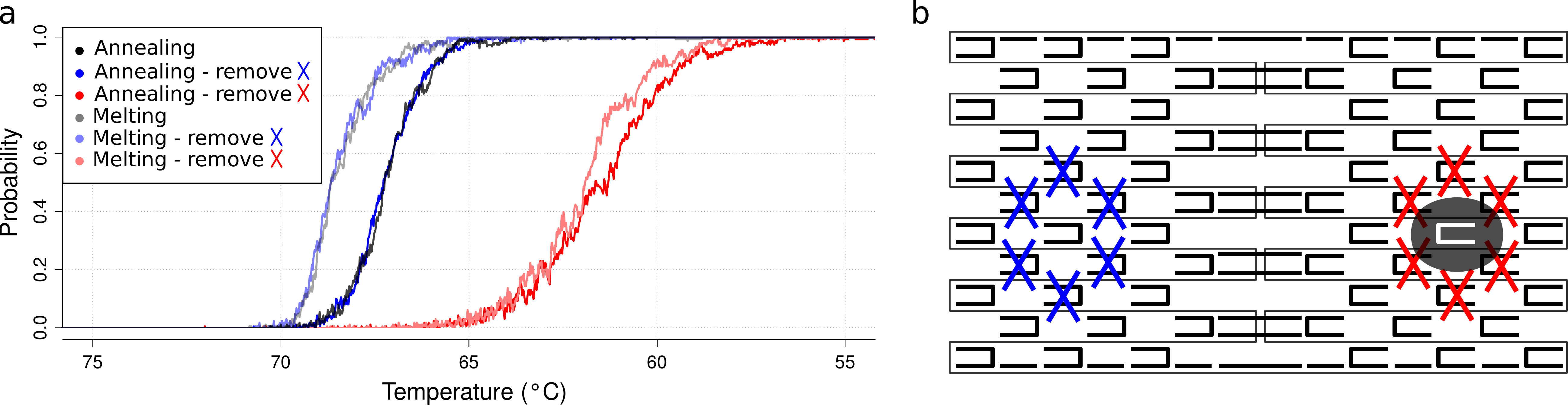}{ 
	Staple insertion in the presence of defects. (a) The probability that staple $X$, highlighted, is fully bound to the scaffold is 
plotted as a function of temperature for simulations in which certain staples are removed from the system (global model,$\gamma=2.5$, $n=2$). 
      Curves are averages over 160 cycles with a temperature ramp of $1 \degreeCpMin$.
	Removing  distant staples, coloured blue in (b), does not significantly change the melting/annealing curves, but
	removing staples around the marked staple (coloured red in (b)) does reduce the incorporation temperature significantly.
	\label{missing_staple_fig}
}

\subsection{Perturbed systems}
We now explore whether the proposed model can capture the consequences of various perturbations to the system. One of the simplest possible perturbations is to reduce the concentration of staple strands \cite{Wei2013}. In Fig. \ref{reduce_concentration}\,(a), we show the effects of reducing the concentration of staples to 2\,nM, retaining the assumption that these staples are in substantial excess over scaffolds (see Fig. \ref{Basic_behaviour}\,(c) for comparable data for a staple concentration of 20\,nM). The clearest effect is the substantial drop in $T_a$, both for the origami a whole and staple $X$ individually; by contrast, the melting transition is largely unchanged from the higher-concentration case. As a result, hysteresis is enhanced:  reduced concentrations slow the rate of hybridization but not melting.
Similar results were reported in an experimental study \cite{Wei2013}. 

There is, however, a more subtle effect at play, which can be most clearly seen if we revert to 20\,nM staple concentrations and compare temperature ramps of $10 \degreeC$ per minute and $0.1 \degreeC$ per minute as shown in Fig. \ref{reduce_concentration}\,(b) and (c). Comparing the two protocols, we see that, on rapid heating, the melting temperature $T_m$ increases by around $1.3 \degreeC$, whereas the annealing temperature $T_a$  decreases by $5.3\degreeC$. Similar behaviour was observed experimentally by Sobczak {\it et al.} \cite{Sobczak2012}, leading those authors to conclude that ``folding rather than unfolding was not in equilibrium''. It is difficult to interpret this conclusion, however, as the question of whether the system is in equilibrium involves the balance or otherwise of folding and unfolding transitions with each other. 

When a system is driven out of equilibrium by fast temperature ramps, one would generally expect the melting temperature to rise and the annealing temperature to fall relative to the equilibrium case, as the system responds to the external driving with some time delay. Within our model we do observe this behaviour, but the shift in the annealing transition is far larger. To understand this asymmetry, note that a major feature of the model is that the rate at which a staple fully binds to the origami is only weakly temperature-dependent (and binding rates of the first domain are temperature-independent) \cite{Morrison1993}. By contrast, the rate at which a two-domain staple unbinds is extremely temperature-dependent, increasing by approximately an order of magnitude following a $2 \degreeC$ increase in temperature (Section \ref{kinetic model}). This increase in melting rate with temperature allows the system to respond more quickly to being  raised rapidly above its equilibrium melting point;  by contrast, annealing rates do not notably increase with decreasing temperature, and so the system responds relatively slowly to being lowered rapidly below its equilibrium melting temperature. 

An alternative perturbation is to remove certain staples from solution; such an experiment was performed by Wei {\it et al.} \cite{Wei2013}. We explore the effect on the incorporation of staple $X$  of removing from solution either: (i) all neighbouring staples or (ii) equivalent staples on the opposite side of the origami (Fig. \ref{missing_staple_fig}). The effects are  clear -- the absence of staples from the local neighbourhood has a substantial influence on the incorporation of staple $X$, whereas more distant staples have very little effect. Local cooperativity, which was also observed by Wei  {\it et al.} \cite{Wei2013}, arises naturally in a model in which interactions between staples are mediated through loops and coaxial stacking. Our model predicts that  hysteresis exhibited by staple $X$ is strongly suppressed if its neighbouring staples are removed, reducing the strength of cooperative effects.  

\section{Summary and Conclusions}
We have presented a domain-level model of DNA origami folding and explored its properties. Our modelling framework is explicitly kinetic and naturally allows for hysteresis, local cooperativity and the natural variability of assembly pathways during origami folding. We define a thermodynamically self-consistent model (our `global' model) that is restricted to planar origami designs. We also define a `local' model of origami folding which is based on the same physical principles but less thermodynamically rigorous. The local model is easier to simulate and can be used for non-planar designs. We  demonstrate that the two approaches give similar results for a simple origami, with small and understandable quantitative differences.

Within a general framework for origami modelling, we consider a specific form for the change in free energy  associated with staple binding. Duplex binding free energies are taken from the SantaLucia parameterization of the nearest-neighbour model of DNA thermodynamics \cite{SantaLucia2004}, with an additional contribution from coaxial stacking of adjacent domains. The free-energy cost of constraining the origami scaffold to form a loop is calculated as $\Delta G^{\rm loop} = - RT \gamma \ln{ \left( {C}/{ \sum_{i \leq m} N_i b_i^2}\right)}$, where $N_i$ is the number of Kuhn lengths of domain type $i$ (ssDNA or dsDNA) present in the loop, $b_i$ is the Kuhn length of  domain type $i$ and $\gamma$ and $C$ are constants. This description is very simple, and is based on approximations that will break down under certain circumstances (for example, the original freely-jointed chain derivation that gives $\gamma = 3/2$ is based on the assumption that there are many Kuhn lengths in the loop). Nonetheless, $\Delta G^{\rm loop}$ captures an important contribution to the cooperativity of origami folding. We show that larger values of the loop exponent $\gamma$ give stronger interactions between staples. Interestingly, we find that the formation of a duplex domain within a loop by hybridization of a single staple domain can increase the free-energy penalty for closing the loop.

We have demonstrated that parameters of the model can be chosen  to give results consistent with experimental data \teoedit{($\gamma=2.5$ and $n=2$)}. Hysteresis in the model is asymmetric (annealing curves are shifted further from the equilibrium transition temperature than melting curves) and a reduction of the concentration of staples primarily influences annealing rates: both observations are consistent with experimental evidence  \cite{Sobczak2012,Wei2013}. Origami folding within the model is naturally locally cooperative:  the omission of selected staples has almost no effect on the insertion of distant staples, as observed experimentally \cite{Wei2013}. This is consistent with the observation that distinct sections of origami can fold independently at different temperatures \cite{Sobczak2012}.  
These physically reasonable findings serve to validate the general framework of the model.

This domain-level model is extremely simple and has considerable potential for improvement. Major simplifications include the lack of an explicit representation of the geometry of the origami. This could have several effects. For example, when an arm of a staple detaches from an otherwise intact origami, the resultant single-stranded scaffold domain will be held in place by the remainder of the origami, reducing its entropy and favouring reformation of the domain in a manner that is only partly captured by our treatment of scaffold loops. More generally, the greater geometrical order imposed by staple binding may contribute to cooperative effects that are  not well described by the model. The calculation of $\Delta G^{\rm loop}$ is  very approximate, and fitting to more detailed data (as was attempted by Arbona {\it et al.} \cite{Arbona2013}) for specific loops may give better quantitative modelling -- although the functional form as it stands is instructive in elucidating the physical effects which are important in staple-staple interactions. We note that  the use of $\gamma = 2.5$, coaxial stacking strength $n=2$ may be compensating for physical effects that are neglected in the model, such as cooperativity mediated by overall scaffold geometry (Arbona {\it et al.} propose a non-specific attractive interaction between helices \cite{Arbona2013}, which could also be incorporated). 

\teoedit{Extending the model to allow interactions other than hybridization between fully complementary staple and scaffold domains would be
desirable. Such interactions include partial hybridization of staples to off-target scaffold domains and scaffold secondary structure. This may prove challenging as such misbonding interactions do not naturally respect the abstraction of DNA
hybridization at the level of domains. Off-target staple-scaffold interactions would have to be remarkably strong to cause significant effects as in the critical temperature range over which folding occurs even the binding of a fully complementary staple domain is transient (an origami with a wider range of scaffold-staple binding strengths may be more prone to staple misbonding). The high temperatures of origami assembly
should also limit the consequences of scaffold secondary structure. Indeed, the MFOLD
software \cite{Zuker2003} predicts that at  65$\,^\circ$C  (a typical origami assembly temperature) there is only one significant secondary structure motif in the scaffold that we use, a 15\,bp hairpin containing a single mismatched base pair.}

Future work will also focus on improving the parameterization of free energies, and exploring the physical consequences of staples with more than two domains or with domains of significantly different lengths. Important open questions include how and why 3D origami folding differs from 2D origami. It remains to be seen whether observed differences, e.g. stronger hysteresis for 3D structures, can be explained with models of the kind proposed here.

\vspace{2mm}

\noindent
\emph{Acknowledgements} FD would like to acknowledge Chris Thachuk for helpful discussions. The authors were supported by Engineering and Physical Sciences Research Council grants EP/G037930/1, EP/P504287/1, a Human Frontier Science Program grant 
GP0030/2013, a Microsoft Research PhD Scholarship (FD), University College Oxford (TEO), the ERC Advanced Grant VERIWARE (MK and FD) and a Royal Society–Wolfson Research Merit Award (AJT).

\bibliography{library}


\appendix


\section{Simulating staples with more than two domains}\label{multidomainstaples}
\subsection{State space}
It is trivial to extend the state space of the model to consider staples with more than two domains. For example, for sections of a scaffold that can bind to three domains, we have the following states $\staple$ (and associated permutations):
\begin{itemize}
  \item	000: no staple bound to  any scaffold domain.
  \item	100: a single staple is bound to the first domain, the second and third domains are empty.
  \item 110: a single staple is bound to the first two domains, the third is empty.
  \item 101: a single staple is bound to the first and third domains, the second is empty.
  \item 120:  distinct staples are bound to the first and second domains, the third is empty.
  \item 111: a single staple is bound to all three domains.
  \item 112: a single staple is bound to the first two domains, a second, distinct staple is bound to the third domain.
  \item 123: distinct staples are bound to all three domains.
\end{itemize}

\subsection{Kinetics of multi-domain staples}
In principle, kinetic models can be constructed in a very similar manner to that described in Section \ref{kinetic model} (Eqs. \ref{Rs00s01}, \ref{Rs01s00}, \ref{Rs11s01} and \ref{internalrate}). Binding of secondary or tertiary domains of a staple introduces looping constraints, as before, and the rates can be defined to take these constraints into account. Duplex and coaxial stacking free energies can be calculated as before -- we discuss additional subtleties associated with the calculation of  $\Delta G^{\rm shape}$, and important topological issues, below. 

\subsection{$\Delta G^{\rm shape}$ in the global model}

  \pictureWithCaptionSingleColumn{b!}{0.99}{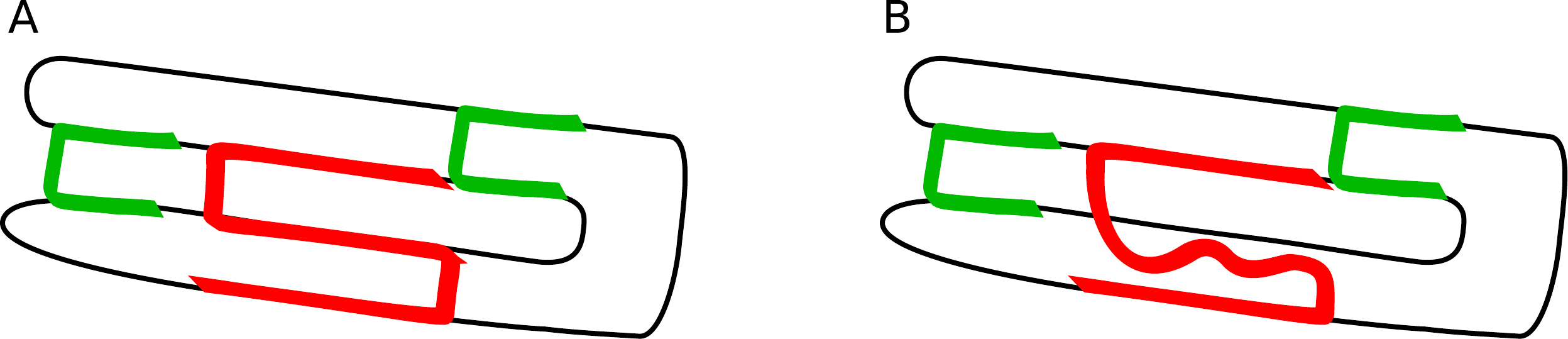}{Three-domain staple in fully bound (A) and semi-bound (B) states. In the case of state B, a planar embedding is not possible (the arcs of the graph cannot be drawn in 2D without them crossing); the global model has to be adapted to deal with this situation.}
  
  \noindent
  
A two-dimensional origami with multi-domain staples can generally be represented as a planar graph (unless parallel crossovers are used). Each staple crossover can be labelled as internal or external based on the intended design. However, intermediate states in which a staple binds by non-adjacent domains (such as $p=101$) are, in general, not planar due to the edge that contains the unbound central staple domain (Fig.\,\ref{multidomain}). 
One solution to this problem would be to forbid states in which a staple binds to non-consecutive domains, such as $p=101$. Such an approach is taken by Arbona {\it et al.} to simplify their model \cite{Arbona2013}. An alternative, less drastic approach might be the following.
\begin{itemize}
\item First, consider only crossovers that link consecutive bound domains of staples (green staples in Fig.\,\ref{multidomain}B), and identify the faces and loops based on the resultant graph $\graph$.
\item Then consider in turn each staple that is bound by non-consecutive domains (red staple in Fig.\,\ref{multidomain}B), assigning loops based on the smallest cycle (see Section \ref{algorithm}) involving a graph constructed using $\graph$ and the staple-link in question.
\end{itemize} 
The result is a thermodynamically well-defined model, in that each bonding configuration can be assigned an unambiguous free energy. 
Note that links formed by the binding of non-consecutive domains of a staple (\cf Fig.~\ref{multidomain}B) are not considered when computing the looping constraints in $\graph$.

\subsection{$\Delta G^{\rm shape}$ in the local model}
As the local model does not require a planar graph, estimates of $\Delta G^{\rm loop}$ can be made directly from a full non-planar graph involving all staples. 

\subsection{Topological considerations}
Multi-domain staples introduce topological assembly problems. For example, the direct transition $p=101 \rightarrow p=111$ (Fig.\,\ref{multidomain}B to Fig.\,\ref{multidomain}A) is impossible, as the central staple domain and the scaffold cannot twist round each other to form a double helix whilst maintaining contact through the exterior domains. The model would have to be modified to take account of such effects. Topological effects should, however, be describable through effective modulations of a domain-level model - for example, to forbid topologically impossible transitions.

\section{The number of faces (loops) in the global model}\label{number of faces}
As noted in Section \ref{shape section}, the graph corresponding to an origami without staples has two faces (one internal, one external). For every two-domain staple that is subsequently added, one more face is formed. Thus, for an origami with $m$ fully-bound two-domain staples, we have $m+2$ faces.

In our global model, faces are interpreted as looping constraints with an associated $\Delta G^{\rm loop}$. Is it problematic that our formalism introduces $m+2$ constraints? A circular origami has, in reality, $m+1$ looping constraints (i.e., an emptly, circular scaffold has a single constraint). So the question is whether the single extra looping constraint in the global model is a problem. To see that it is not, consider the following:
\begin{itemize}
\item whether the total number of loop constraints is $m+2$ or $m+1$, the effect of adding an extra staple is to create an additional constraint;
\item the effect of the large external face on $\Delta G^0_{s,s^\prime}$ is always small. In the initial state, the external face of the origami tile shown in Fig. \ref{layout} is associated with a loop of $2646$ nucleotides ($E[\distance^2]=2858 \mbox{ nm}^{2}$). As staples are added, the loop remains large, although in the final state it is somewhat smaller, consisting of $66$ duplex domains and $10$ staple crossovers ($E[\distance^2]\approx 1986 \mbox{ nm}^{2}$). Even for $\gamma = 2.5$, $\Delta G^{\rm loop}_{37 \degreeC}$ of the large external loop only changes by $0.56 $ kcal/mol between the limiting cases of the free scaffold and fully-folded state. The change in $\Delta G^{\rm shape}_{s,s^\prime}$ due to the external face between any two intermediate states $s$ and $s^\prime$ that can interconvert is therefore always very small. Given the  approximate nature of the model, these small differences are not significant. 
\end{itemize}
A possible alternative approach would be to ignore the large external loop in calculating $\Delta G^{\rm shape}$ -- complicating the model in this way seems unnecessary at this stage,  however.

  \begin{table*}
	\setlength{\tabcolsep}{6pt}
	\begin{tabular}{ll|l|l|l}
State	&	Face		&	Edges									& 	Weight 								& $E[\distance^2]$ /nm$^2$	\\	\hline
$\stateA$ & $ \face_{1}$	&   $   e_{1}, {\underline e}_{2}, c_{1}, {\underline e}_{9}, e_{10} $		&   $2 \times (xl_{ds})^2 + (2xl_{ss}/\lambda_{ss} + 1) \times \lambda^2_{ss} $    & 97.0	\\
	  & $ \face_{2}$	&   $ e_{3}, {\underline e}_{4}, e_{5}, e_{6}, e_{7}, e_{8}, c_{1} $		&   $1 \times (xl_{ds})^2 + (5xl_{ss}/\lambda_{ss} + 1) \times \lambda^2_{ss} $    & 119.2	\\	
	  & $ \face_{3}$	&   $ e_{1}, {\underline e}_{2}, e_{3}, {\underline e}_{4}, e_{5}, e_{6}, e_{7}, e_{8}, {\underline e}_{9}, e_{10} $	&   $3 \times (xl_{ds})^2 + (7xl_{ss}/\lambda_{ss} + 0) \times \lambda^2_{ss} $    & 209.7	\\ \hline
$\stateB$ & $ \face_{1}$	&   $ e_{1}, {\underline e}_{2}, c_{1}, {\underline e}_{9}, e_{10} $		&   $2 \times (xl_{ds})^2 + (2xl_{ss}/\lambda_{ss} + 1) \times \lambda^2_{ss} $    & 97.0	\\	
	  & $ \face_{2}$	&   $ e_{3}, {\underline e}_{4}, c_{2}, {\underline e}_{7}, e_{8}, c_{1} $		&   $2 \times (xl_{ds})^2 + (2xl_{ss}/\lambda_{ss} + 2) \times \lambda^2_{ss} $    & 100.2	\\
	  & $ \face_{3}$	&   $ e_{5}, e_{6}, c_{2} $								&   $0 \times (xl_{ds})^2 + (2xl_{ss}/\lambda_{ss} + 1) \times \lambda^2_{ss} $    & 37.8	\\
	  & $ \face_{4}$	&   $ e_{1}, {\underline e}_{2}, e_{3}, {\underline e}_{4}, e_{5}, e_{6}, {\underline e}_{7}, e_{8}, {\underline e}_{9}, e_{10} $	&   $4 \times (xl_{ds})^2 + (6xl_{ss}/\lambda_{ss} + 0) \times \lambda^2_{ss} $  & 
222.0		\\ \hline
$\stateC$ & $ \face_{1}$	&   $ e_{1}, {\underline e}_{2}, c_{1}, {\underline e}_{9}, e_{10} $		&   $2 \times (xl_{ds})^2 + (2xl_{ss}/\lambda_{ss} + 1) \times \lambda^2_{ss} $    & 97.0	\\	
	  & $ \face_{2}$	&   $ e_{3}, e_{4}, e_{5}, e_{6}, e_{7}, e_{8}, c_{1} $				&   $0 \times (xl_{ds})^2 + (6xl_{ss}/\lambda_{ss} + 1) \times \lambda^2_{ss} $    & 106.9	\\
	  & $ \face_{3}$	&   $ e_{1}, {\underline e}_{2}, e_{3}, e_{4}, e_{5}, e_{6}, e_{7}, e_{8}, {\underline e}_{9}, e_{10} $	&   $2 \times (xl_{ds})^2 + (8xl_{ss}/\lambda_{ss} + 0) \times \lambda^2_{ss} $ & 197.3   	\\	\hline
$\stateD$ & $ \face_{1}$	&   $ e_{1}, {\underline e}_{2}, c_{1}, {\underline e}_{9}, e_{10} $		&   $2 \times (xl_{ds})^2 + (2xl_{ss}/\lambda_{ss} + 1) \times \lambda^2_{ss} $    & 97.0	\\	
	  & $ \face_{2}$	&   $ e_{3}, {\underline e}_{4}, e_{5}, e_{6}, {\underline e}_{7}, e_{8}, c_{1} $	&   $2 \times (xl_{ds})^2 + (4xl_{ss}/\lambda_{ss} + 1) \times \lambda^2_{ss} $    & 131.5	\\
	  & $ \face_{3}$	&   $ e_{1}, {\underline e}_{2}, e_{3}, {\underline e}_{4}, e_{5}, e_{6}, {\underline e}_{7}, e_{8}, {\underline e}_{9}, e_{10} $	&   $4 \times (xl_{ds})^2 + (6xl_{ss}/\lambda_{ss} + 0) \times \lambda^2_{ss} $ & 222.0	\\ \hline
	\end{tabular}
	\caption{
	Weights as given by the global model for the faces of the system depicted in Fig. \ref{rates}. For simplicity, we take all scaffold domains to have the same length of 16\,nt, with contour length $L=16 \times 0.34\,$nm in the single-stranded state. Double-stranded edges are underlined. 
	\label{tableFaces}
	}
\end{table*}

\section{Examples of loop and rate calculations} \label{computing rates}
  \pictureWithCaptionSingleColumn{h}{0.99}{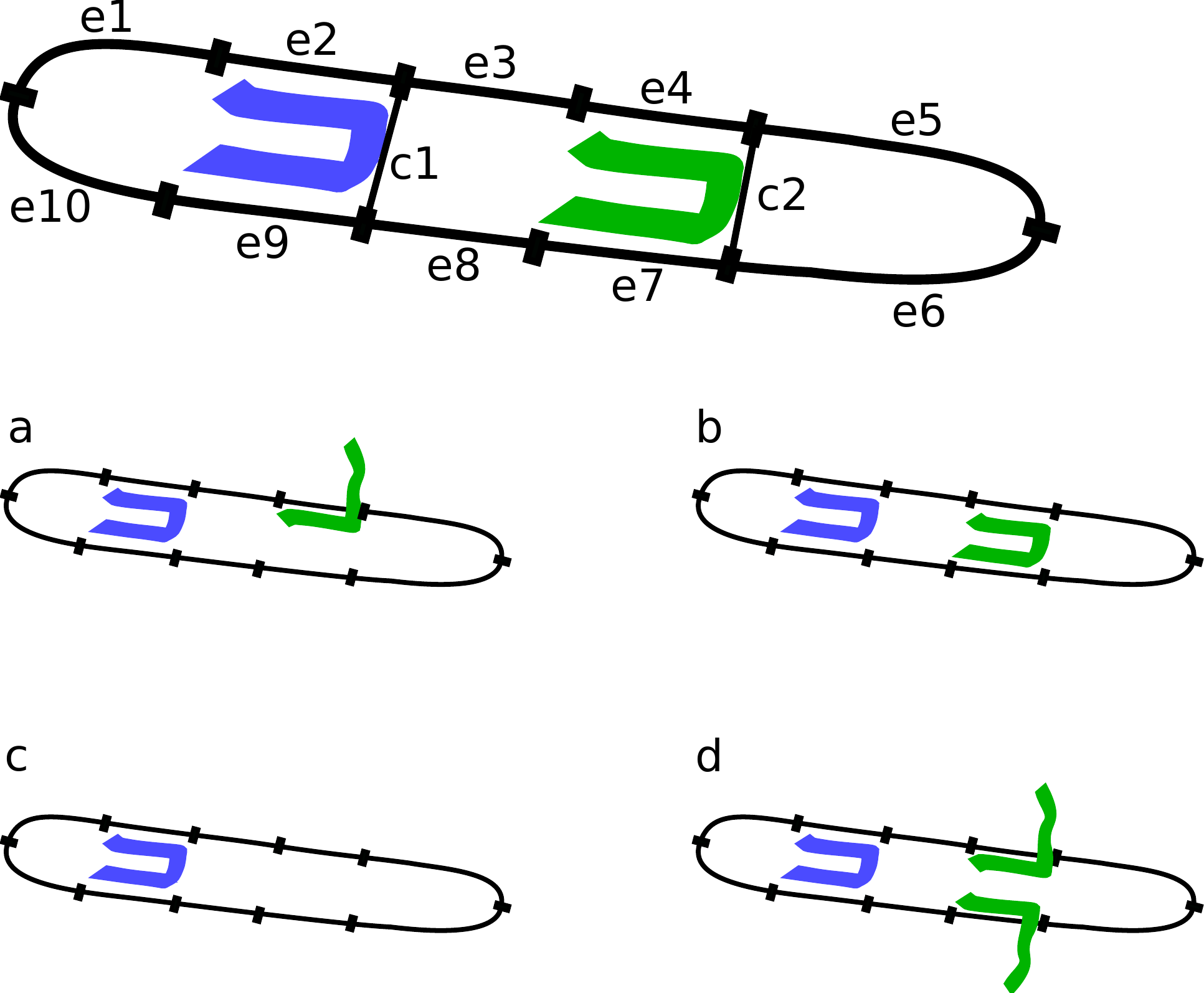}{Graph representation with labelled edges of an  origami in four  partly-folded states ($\stateA - \stateD$). The faces are enumerated in Table \ref{tableFaces}. 
  }

  Fig. \ref{rates} shows a partly-folded origami in a variety of states: we discuss the calculation of the term $\Delta G^{\rm {shape}}$ for  transitions between these states.
  
  Firstly, in both the global and local models, initial binding from solution and unbinding of a second staple domain do not require estimation of $\Delta G^{\rm shape}$. Transitions $\rate (\stateB,\stateA)$ and  
    $\rate(\stateC,\stateA)$ fall into these categories. Following Eq.\,\ref{Rs00s01} and Eq.\,\ref{Rs11s01}, we find 
  \begin{align}
  \rate (\stateC,\stateA) & = k_+ [{\rm staple}] 
  \intertext{and}
  \rate (\stateB,\stateA) & = k_+ \exp \left(\Delta G^{0\,{\rm duplex}}_{\stateA,\stateB}/RT \right) \times {\rm M},
  \end{align}
 in which $\Delta G^{0\,{\rm duplex}}_{\stateA,\stateB}$ is the estimated standard free-energy change of formation of the duplex in question and $[{\rm staple}]$ is the staple concentration (there are no coaxial stacking changes to consider in this case).

  In the global model, all other transitions require estimates of $\Delta G^{\rm shape}$. In Table \ref{tableFaces}, we identify the relevant faces and associated loop costs (a subset of loop costs are also relevant to the local model). We recall that loop costs are determined by $E[\distance^2]$, which is in turn given by summing over the weights of individual edges in a loop.
    These edges contribute (Eq.\,\ref{W(e)}):
    \begin{align}
    \weight(c_{i}) & =  \lambda_{ss}^2  = (1.8\,{\rm nm})^2	\quad \mbox{for a crossover;} 	  	\\
    \weight(e_{i}) & =
      \begin{dcases}
	(xl_{ds})^{2} = (0.34\,{\rm nm} \times x)^2 		& \mbox{for $x$-bp dsDNA} 	  		\\
	 xl_{ss}\lambda_{ss} = 1.08\,{\rm nm}^2 \times   x 	& \mbox{for $x$-nt ssDNA} 	  	\\
      \end{dcases}	    
    \end{align}
   
  These expressions are used in Table \ref{tableFaces} to calculate $E[\distance^2]$ for the various faces in the global model.

\subsection{Global model}
In the global model the shape term is computed as (\eqn \ref{result})
  \begin{align}
   \frac{\Delta G^{\rm shape}_{s,s^{\prime}}}{-RT\gamma}	&	= \sum_{\faces(\state^{\prime})} \ln \frac{C}{\weight(\face_{i})} -  \sum_{\faces(\state)} \ln \frac{C}{\weight(\face_{i})}
  \end{align}
where $ \faces(\state)$ are the faces in the embedding of $\graph(\state)$ and $\weight(\face_{i})$ is the weight of a face.
{Taking $\gamma=2.5$ (and $C_{2.5}=2.8 \times 10^{-18} \mbox{m}^{2}$), we find at $T=37.0 \degreeC$
\begin{align}
  \Delta G^{\rm shape}_{\stateA,\stateB}  & = 3.828	\text{ kcal/mol,} 				\\
  \Delta G^{\rm shape}_{\stateA,\stateC}  & = -0.261 	\text{ kcal/mol, and}  				\\
  \Delta G^{\rm shape}_{\stateD,\stateA}  & = -0.239 	\text{ kcal/mol.}
\end{align}
 using  Combining these specific values with Eq.\,\ref{Rs01s00} and Eq.\,\ref{internalrate}. we find 
\begin{align}
\rate(\stateA,\stateC) & = 0.6546 \cdot k_+ \exp \left(\Delta G^{0\,{\rm duplex}}_{\stateC,\stateA}/RT \right) \times {\rm M}, 	\\
\rate(\stateD,\stateA) & =  0.6782 \cdot k_+ \exp \left(\Delta G^{0\,{\rm duplex}}_{\stateA,\stateD}/RT \right) \times {\rm M}, 	\\
\rate(\stateA,\stateB) & = 1.998 \cdot 10^{-3} \times k_+ \times {\rm M}.							
\end{align}
It is clear that the unbinding rate for a half-bound staple (transition $\stateA \rightarrow \stateC$) depends on the state of the origami, while the unbinding rate for a domain of a fully-bound staple ($\stateC \rightarrow \stateA $) does not. This is a direct result of the choice of rate constants recorded in Equations \ref{Rs00s01} and \ref{Rs11s01}. Different approaches are possible without violating thermodynamic consistency of the model.

\subsection{Local model}

For transitions $\rate(\stateA,\stateC)$ and $\rate(\stateD,\stateA)$, the local model uses $\Delta G^{\rm shape}=0$ as no loops are formed
during these transitions. Thus
 \begin{equation}
\begin{array}{c}
\rate(\stateA,\stateC) =  k_+ \exp \left(\Delta G^{0\,{\rm duplex}}_{\stateC,\stateA}/RT \right) \times {\rm M}, \\
  \\
\rate(\stateD,\stateA)  = k_+ \exp \left(\Delta G^{0\,{\rm duplex}}_{\stateA,\stateD}/RT \right) \times {\rm M}. \\
\end{array}
\end{equation}
In case of the transition $\rate(\stateA,\stateB)$, a new loop is formed. Thus we must find the cycle in $\stateB$ containing $c_2$ that minimizes $E[\distance^2]$ -- this is the loop consisting of edges
\begin{align}
   e_{5}, e_{6}, c_{2}.
\end{align}
Using the value for $E[\distance^2]$ for this loop tabulated in (Table \ref{tableFaces}),  with Eq.\,\ref{internalrate}, Eq.\,\ref{dGshape} and Eq.\,\ref{result}, we find
\begin{equation}
\rate(\stateA,\stateB) = 1.494 \cdot 10^{-3} \times k_+ \times {\rm M}
\end{equation}

\section{Justification of the form of $\Delta G^{\rm loop}$}\label{DG justification}
 To justify our treatment of the thermodynamic cost of forming a single loop in our model, we consider the thermodynamic cycle shown in Fig. \ref{Theory1}. We consider the binding of two domains to a longer strand, both when they are connected to form a single staple and when they are separate strands. 

  \pictureWithCaptionSingleColumn{h}{0.99}{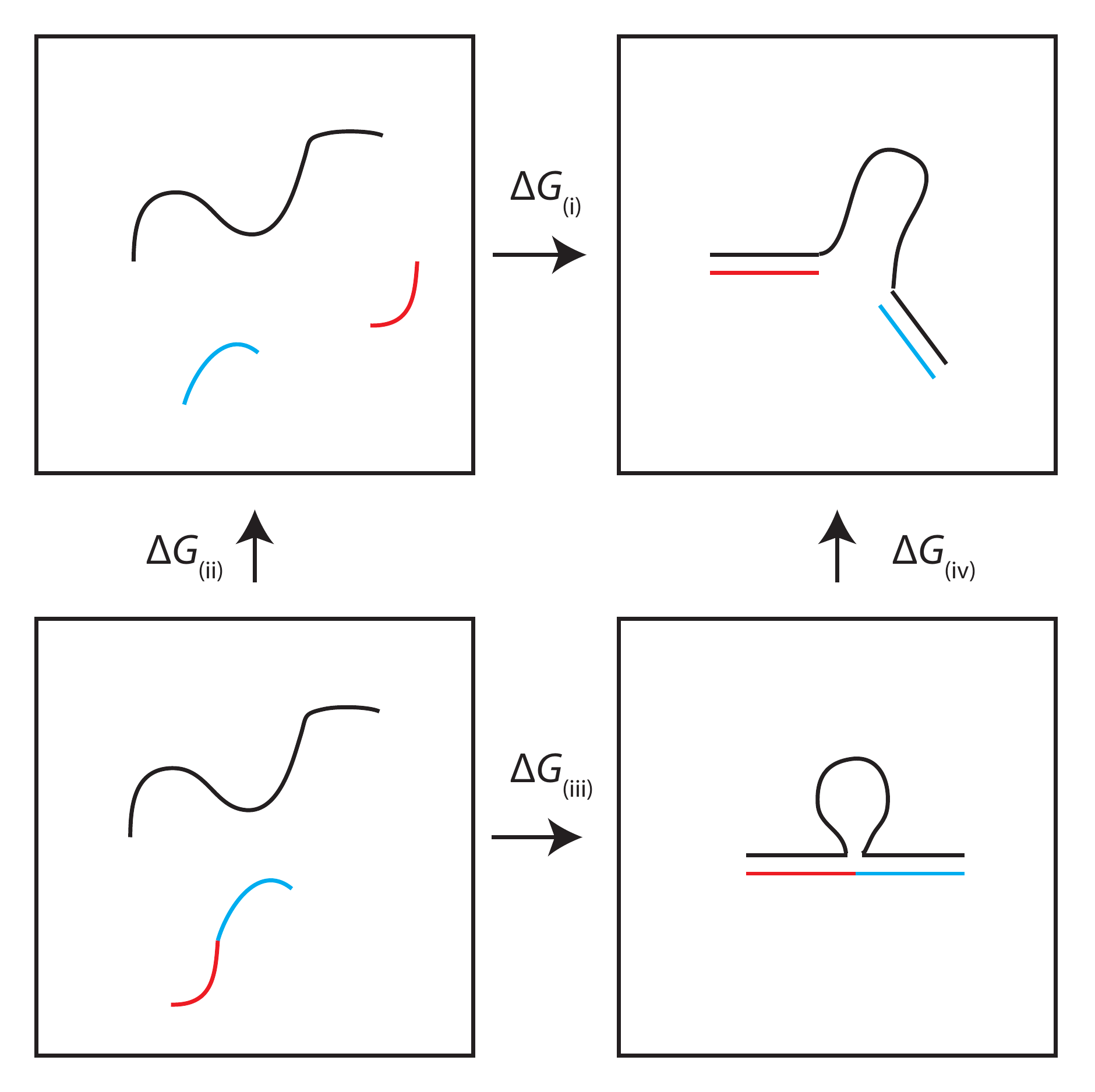}
{Thermodynamic cycle. We consider two domains, $a$ (red) and $b$ (blue), that can bind to a scaffold. We consider the thermodynamics of binding when the two are connected  and when they are separate. Estimates of the free energies of transitions (i), (ii) and (iv) are used to estimate that of transition (iii).
}

We take the entire system to be contained within an arbitrary volume $v$. The free energy change associated with transition (i) is simply the sum of those for the formation of two independent duplexes:
\begin{equation}
\Delta G_{(i)} = \Delta G^0_a + \Delta G^0_b + 2RT \ln (v / v^0),
\label{Gi}
\end{equation}
where the first two terms are standard free energies of formation, and the third corrects for the fact that standard free energies are defined at 1\,M concentration of reactants \cite{Ouldridge_bulk_2010}.

\pictureWithCaption{!}{0.99}{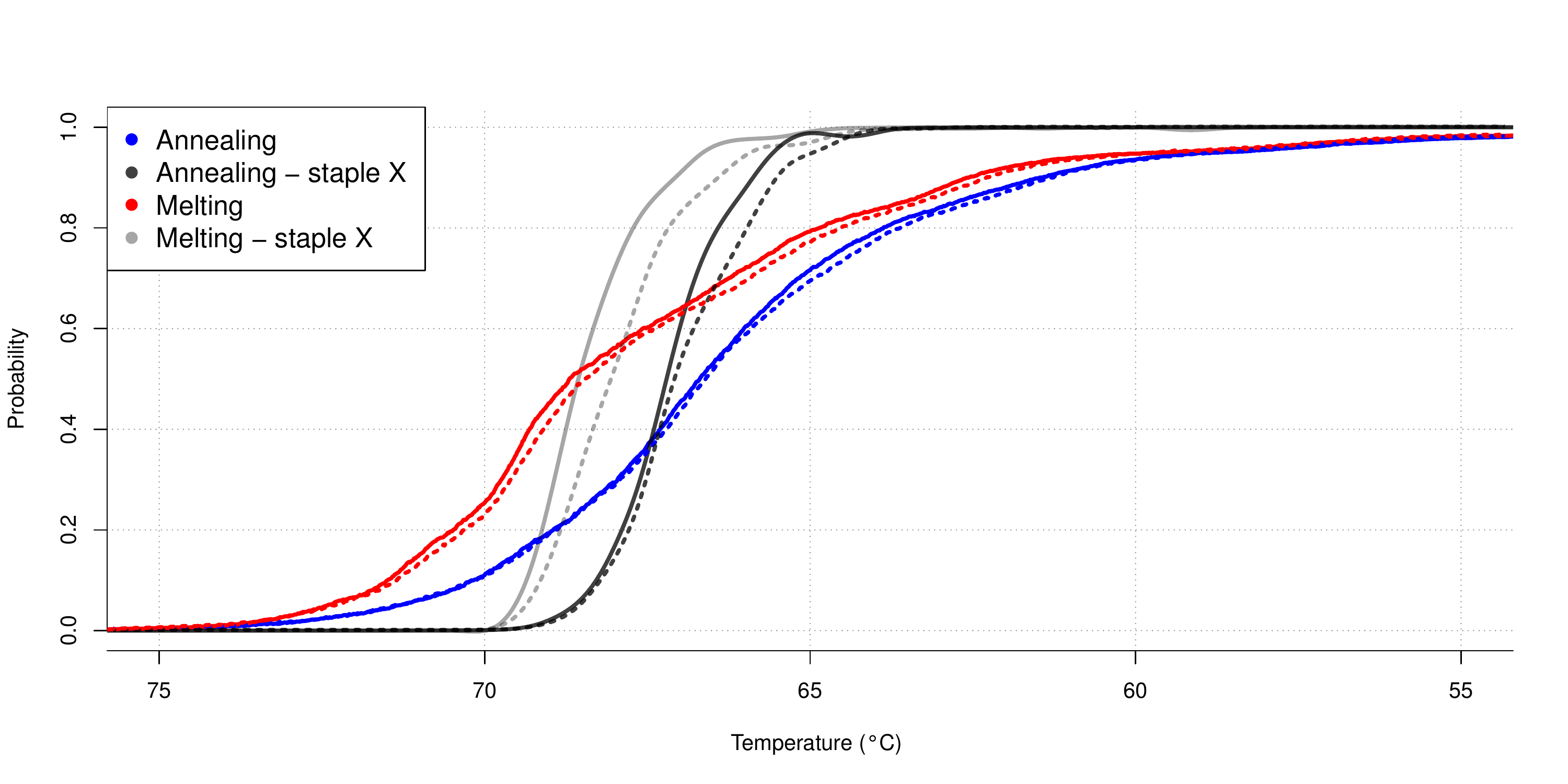}{Staple binding as a function of temperature during cooling and heating of the simple origami for coaxial stacking constant $n=2$ and  loop exponent $\gamma = 2.5$, staple concentration $20$ nM 
and temperature gradient $|dT/dt|=1.0 \degreeCpMin $. Full lines: global model; dashed lines: local model. Red/blue lines indicates behaviour averaged over all staples; black/grey are predictions for a specific staple X (cf. Fig. \ref{reduce_concentration}) -- these data for X were smoothed using a cubic spline.   }

\begin{table}
	\setlength{\tabcolsep}{6pt}
	\begin{tabular}{l|rrr}
		\multicolumn{3}{l}{\quad \quad $T_{a}$/$\degreeC$}		\\
		\multicolumn{1}{l}{} 							\\
		Coaxial stacking	     &	\multicolumn{3}{c}{ Local model - $\gamma$ }	\\
		 $n$ & $1.5$ & $2.5$    & $3.5$ \\
		\hline
		0                  & 64.2  & 62.7     & 61.4   \\
		1.0                & 66.0  & 64.7     & 63.6  \\
		2.0                & 67.5  & 66.3     & 65.2  \\
		3.0                & 68.7  & 67.5     & 66.6  \\
		\multicolumn{4}{l}{} 							\\
		\multicolumn{4}{l}{  \quad \quad  Hysteresis/$\degreeC$ } 	\\
		\multicolumn{1}{l}{} 							\\
		\hline
		0                  & 0.2   & 0.2     &  0.2   \\
		1.0                & 0.5   & 0.5     &  0.4   \\
		2.0                & 0.9   & 0.8     &  1.0   \\
		3.0                & 1.5   & 1.3     &  1.5 \\
		\multicolumn{4}{l}{} 							\\
		\multicolumn{4}{l}{ \quad \quad $\Delta T_{a}$/$\degreeC$ } 		\\
		\multicolumn{1}{l}{} 							\\
		\hline
		0                  & 9.5   & 9.8     & 10.5   \\
		1.0                & 7.1   & 7.5     & 8.0   \\
		2.0                & 5.5   & 5.8     & 6.4   \\
		3.0                & 4.7   & 4.8     & 5.5   
	\end{tabular}
	\caption{
	Predictions of the local model when modified to remove effects of loop-mediated cooperativity (see text). Data is comparable to that presented in Table \ref{table1}.	
			\label{tableNoCooperativity}	
	}
\end{table}

The free energy change associated with transition (ii) includes $\Delta G_{\rm break}$, the cost of  breaking the DNA backbone (independent of $v$). It also includes the entropy gain associated with allowing the two separate halves to explore  volume $v$ independently which depends on the probability, $P^{r_{c}}_v$, that two unconnected halves spontaneously come within an  interaction distance, $r_{c}$, that is comparable to the length of a backbone link:
\begin{equation}
\Delta G_{(ii)} = \Delta G_{\rm break} + RT \ln (P^{r_{c}}_v),
\end{equation}
Transition (iv) also involves the breaking of the DNA backbone, but this time the freedom gained is not the ability to move around the whole of $v$, but rather the freedom for the scaffold to explore conformations that were inaccessible when it was constrained to form a loop. Thus, instead of $P^{r_{c}}_v$, we obtain $P^{r_{c}}_{\rm loop}$, the probability that the ends of the two halves connected by the loop spontaneously come within $r_{c}$ when they are not directly connected:
\begin{equation}
\Delta G_{(iv)} = \Delta G_{\rm break} + RT \ln (P^{r_{c}}_{\rm loop}).
\end{equation}
By inspection of the thermodynamic cycle:
\begin{equation}
\Delta G_{(iii)} = \Delta G_{(ii)} + \Delta G_{(i)} - \Delta G_{(iv)},
\end{equation} 
and thus
\begin{equation}
\begin{array}{c}
\Delta G_{(iii)} = \Delta G^0_a + \Delta G^0_b 
+ RT \ln  \left(\frac{v^2 P^{r_{c}}_v} { (v^0)^2 P^{r_{c}}_{\rm loop}} \right). 
\end{array}
\end{equation}
$P^{r_{c}}_v$  scales as $1/v$, and so we can replace $P^{r_{c}}_v$ with $\frac{v^0}{v}P^{r_{c}}_{v^0}$ to cancel one of the factors of $v / v^0$ inside the logarithm. Overall, reaction (iii) is a bimolecular association, and therefore the free-energy change in a volume $v$ can be converted into a standard free energy by adding a term $RT\ln(v^0/v)$, cancelling the second factor of $v / v^0$  and giving
\begin{equation}
\begin{array}{c}
\Delta G^0_{(iii)} = \Delta G^0_a + \Delta G^0_b +  RT \ln   \left(\frac{P^{r_{c}}_{v^0}} { P^{r_{c}}_{\rm loop}} \right). 
\end{array}
\end{equation}
We can compare with Eq.\,\ref{dG_total} which also describes a change of state in which a staple fully binds to the origami. $ \Delta G^0_a + \Delta G^0_b$ corresponds to the $\Delta G^{0, \rm duplex}$ contribution, and $ RT \ln   \left({P^{r_{c}}_{v^0}} /{ P^{r_{c}}_{\rm loop}} \right) $ to $\Delta G^{\rm shape}$ as required. Any changes in $\Delta G^{\rm stack}$ due to the presence of adjacent duplex domains could be incorporated into this derivation without influencing its outcome.

\section{Additional data}\label{additional data}
\subsection{Effects of changing $\gamma$ that are not dependent on cooperativity between loops}
In Section \ref{results}, we argue that the increasing $\gamma$ would tend to increase transition widths in the absence of cooperative interactions between staple loops. To quantify this effect, we simulate a system from which the effects of loop-mediated cooperativity have been artificially removed. This change is simple for the local model  -- we calculate $\Delta G^{\rm shape}$ for the binding of a staple to an otherwise empty origami, and use this value throughout the simulation regardless of the state of the origami (we note that in this limit, the local and global models are extremely similar and the local model is also thermodynamically well-defined). The results presented in Table \ref{tableNoCooperativity} show that,  in the absence of cooperative interactions between loops, $\Delta T_{a}$ increases with $\gamma$ as expected. 

\subsection{Comparison of local and global models }}

 Fig. \ref{comparisonLoopDistance} shows the predictions of both global and local models. Melting and annealing curves are plotted for coaxial stacking parameter $n=2$,  loop exponent $\gamma = 2.5$, 
 staple concentration $20$ nM and temperature gradient $|dT/dt|=1.0 \degreeCpMin $.
 The predictions of the two models for the average probabilities of staple incorporation are  very similar, consistent with data presented in Table  \ref{table1}. The response for the single staple X shows more variation between the two models.

\end{document}